\documentclass[journal,draftcls,onecolumn,12pt]{IEEEtran}

\usepackage[utf8x]{inputenc}
\usepackage[T1]{fontenc}
\usepackage{eurosym}
\usepackage{color}
\usepackage{bbold}
\usepackage{graphicx}
\usepackage[parfill]{parskip} 
\usepackage{array}
\usepackage{amsmath}
\usepackage{subfigure}

\usepackage{cite}
\usepackage{url}
\let\Pr\relax
\DeclareMathOperator{\Pr}{\mathbb{P}}

\newcommand{\mvtoc}{\alpha} 
\newcommand{\mctov}{\beta} 
\newcommand{\mvtocN}{\tilde{\alpha}} 
\newcommand{\mctovN}{\tilde{\beta}} 
\newcommand{\minit}{\beta_0} 
\newcommand{\Phiv}{\Phi_v} 
\newcommand{\Phic}{\Phi_c} 
\newcommand{\Phia}{\Phi_a} 
\newcommand{\Xseq}{\mathbf{x}^n} 
\newcommand{\Yseq}{\mathbf{y}^n} 
\newcommand{\XseqE}{\mathbf{\hat{x}}^n} 
\newcommand{\Pv}{P^{(\ell)}} 
\newcommand{\PvV}{\mathbf{P}^{(\ell)}} 
\newcommand{\Pc}{Q^{(\ell)}} 
\newcommand{\PvN}{\tilde{P}^{(\ell)}} 
\newcommand{\PvNV}{\mathbf{\tilde{P}}^{(\ell)}} 
\newcommand{\PcN}{\tilde{Q}^{(\ell)}} 
\newcommand{\Pe}{\tilde{p}_e^{(\ell)}} 
\newcommand{\PeN}{\tilde{p}_{e,n}^{(\ell)}} 
\newcommand{\Pinit}{P^{(0)}} 
\newcommand{\thgb}{b_{\ell}} 
\newcommand{\thgbZ}{b_{0,\ell}} 
\newcommand{\thgbO}{b_{1,\ell}} 
\newcommand{\epp}{\epsilon_{01}} 
\newcommand{\epn}{\epsilon_{10}} 
\newcommand{\M}{\mathcal{M}} 
\newcommand{\Q}{\mathcal{Q}} 
\newcommand{\alp}{\gamma} 
\newcommand{\off}{\lambda} 
\newcommand{\offp}{\lambda_0} 
\newcommand{\offn}{\lambda_1} 
\newcommand{\Pt}{\Pi}
\newcommand{\mi}{m} 

\usepackage[acronym]{glossaries}
\newacronym{iid}{i.i.d.}{independent and identically distributed} 

\newcommand{\fixme}[2]{\ifx&#2&{\color{red}#1}\else{\color{red}FIXME\{}#1{\color{red}\}}\footnote{{\color{red}#2}}\PackageWarning{Fixme}{#1: #2}\fi}

\newcommand{\fig}[4]{ \begin{figure}[#4]
  \centering
   \includegraphics[width=#3\columnwidth]{Figures/#1}
   \caption{#2}\label{fig:#1}
  \end{figure}
}

\title{Noisy Density Evolution With Asymmetric Deviation Models}

\author{Elsa Dupraz$^1$ and François Leduc-Primeau$^2$ \\
\small $^1$ IMT Atlantique, Lab-STICC, UMR CNRS 6285, France. \\ 
\small $^2$ Department of Electrical Engineering, Polytechnique Montreal, QC, Canada.}

\begin{document}

\maketitle

\begin{abstract}
This paper considers low-density parity-check (LDPC) decoders affected by deviations introduced by the electronic device on which the decoder is implemented. 
Noisy density evolution (DE) that allows to theoretically study the performance of these LDPC decoders can only consider symmetric deviation models due to the all-zero codeword assumption. 
A novel DE method is proposed that admits the use of asymmetric deviation models, thus widening the range of faulty implementations that can be analyzed.
DE equations are provided for three noisy decoders: belief propagation, Gallager~B, and quantized min-sum (MS). 
Simulation results confirm that the proposed DE accurately predicts the performance of LDPC decoders with asymmetric deviations.
Furthermore, asymmetric versions of the Gallager~B and MS decoders are proposed to compensate the effect of asymmetric deviations.
The parameters of these decoders are then optimized using the proposed DE, leading to better ensemble thresholds and improved finite-length performance in the presence of asymmetric deviations.
\end{abstract}

\section{Introduction}
\IEEEPARstart{I}{n} most applications of error-correction codes, the implementation complexity of the decoder is a primary concern, and energy consumption is an important factor limiting the performance of the codes.
Unfortunately, it is becoming increasingly difficult to improve the energy efficiency of integrated circuits while maintaining the abstraction that memory and computation circuits behave deterministically\cite{gupta:2013}. 
A growing body of work is therefore being devoted to the study of digital systems built out of unreliable circuit components. \textcolor{black}{For instance,~\cite{yang2017computing} considers fault-tolerant linear computing,~\cite{yang2016fault} addresses logistic regression on unreliable hardware, ~\cite{dupraz2019binary} considers noisy binary recursive estimation, and~\cite{hacene2019training} proposes to train deep neural networks for robustness to hardware faults.}
Error-correction codes are closely related to the development of such systems, first because of the interest of developing more energy-efficient decoder implementations, but also because they can be used within a computing system to restore the fully-reliable operation abstraction when it might be required.

There has thus been significant interest in studying the operation of low-density parity-check (LDPC) decoders in contexts where values stored in memory and/or the result of computations can be affected by errors, or \emph{deviations}\cite{varshney2011performance, leduc-primeau:2012, al2014fault, sundararajan2014noisy, Huang2014Gallager, ngassa2015density, balatsoukas2014density, leduc2018modeling, Dupraz15Com}.
%
Deviations in memory or computation circuits result from the difficulty or impossibility of predicting the variations in the physical properties of the circuit that occur at the time of fabrication or during operation of the system. 
Deviations can be prevented by operating the system based on the worst-case conditions, but this can be very costly in terms of energy consumption or performance. Instead, deviations can be modeled and their impact taken into account.

Density evolution (DE)~\cite{richardson2001capacity,richardson2001design} is a powerful tool for the performance analysis of LDPC codes. 
DE consists of calculating the successive probability density functions of the messages exchanged in the iterative LDPC decoder, which permits to evaluate the decoder error probability under given channel conditions. 
For a given code degree distribution, DE thus allows to predict the \emph{code ensemble threshold} as the worst channel parameter that allows for a vanishing decoding error probability over the ensemble of codes that follow the considered degree distribution.
The threshold is evaluated assuming that the codeword length tends to infinity. 
\textcolor{black}{In this paper, we are interested in analyzing the performance of decoders affected by deviations that can be asymmetric, in the sense that the probability that a logic 0 deviates to a logic 1 is not necessarily the same as the probability that a logic 1 deviates to a logic 0. When such asymmetric deviations are applied to the messages exchanged in an LDPC decoder, the symmetry of the decoder in the sense of \cite{richardson2001capacity} no longer holds.}
Standard DE~\cite{richardson2001capacity,richardson2001design} requires symmetry assumptions on the channel, the variable node (VN) mapping, and the check node (CN) mapping of the decoder.  Under the symmetry assumptions, it can be shown that the decoder error probability does not depend on the transmitted codeword.
This allows to calculate the successive message probability density functions under the assumption that the all-zero codeword was transmitted, and greatly simplifies the calculation. 
The symmetry assumptions were removed in~\cite{wang2005density,chen2009equivalence}, which consider non-symmetric channels, VN, and CN mappings. In~\cite{wang2005density,chen2009equivalence}, the all-zero codeword assumption cannot be considered and the successive message probabilities are conditioned on the codeword bits. 

The literature on faulty LDPC decoders\cite{varshney2011performance, leduc-primeau:2012, al2014fault, sundararajan2014noisy, Huang2014Gallager, ngassa2015density, balatsoukas2014density, leduc2018modeling, Dupraz15Com} assumes that deviations are introduced by the hardware in the VN and CN mappings. 
In this setup, noisy DE was introduced in~\cite{varshney2011performance} and latter considered for the analysis of bit-flipping decoders~\cite{al2014fault,sundararajan2014noisy}, Gallager~B decoders~\cite{leduc-primeau:2012,Huang2014Gallager}, Min-Sum decoders~\cite{ngassa2015density,balatsoukas2014density,leduc2018modeling}, and finite-alphabet iterative decoders (FAIDs)~\cite{Dupraz15Com}.  
Most of these works consider symmetric deviation models in addition to symmetric channels, VN, and CN mappings, and their DE analyses consider the all-zero codeword assumption.
The only exception is~\cite{Huang2014Gallager}, which considers non-symmetric deviation models in the Gallager~B decoder. However, the DE expressions obtained in~\cite{Huang2014Gallager} depend on the codeword weight, which makes them difficult to manipulate. The theoretical part of~\cite{Huang2014Gallager} proposes approximations for the DE expressions. However, the proposed approximations only apply to the Gallager~B decoder with hard-decision messages. 

In this paper, we consider noisy DE with asymmetric deviation models and without the all-zero codeword assumption. The DE expressions we derive however do not depend on the codeword weight as in~\cite{Huang2014Gallager}. 
Instead, these expressions are conditioned on the codeword bit $0$ or $1$ at the VN, following the approach of~\cite{wang2005density} for standard (non-faulty) DE. 
This approach leads to DE expressions that are much simpler than in~\cite{Huang2014Gallager}, and that can be applied to any type of LDPC decoder. 
We introduce the general noisy DE method under asymmetric deviation models, and then apply this method to three particular LDPC decoders: Belief Propagation (BP), Gallager~B, and quantized Min-Sum (MS) decoder.
For each considered decoder, we provide the DE equations under specific deviation models. 
In addition, in order to compensate asymmetric deviations in the decoder, we propose to introduce asymmetric parameters in the Gallager~B decoder and in the quantized MS decoder. For the Gallager~B decoder, we assume that the decision threshold at VNs is asymmetric and depends on the sign of the channel output. For the quantized MS decoder, we assume that the offset parameter and the scaling parameter are asymmetric, and depend on the sign of the CN messages and on the sign of the channel outputs, respectively. For the two decoders, we show that asymmetric decoder parameters allow to improve the decoders threshold as well as their finite-length performance. 
\textcolor{black}{Finally, the main contributions of the paper can be summarized as follows:
\begin{enumerate}
 \item We extend the asymmetric DE analysis without deviations of~\cite{wang2005density}  in order to capture the effect of asymmetric deviations onto the LDPC decoders performance. For particular asymmetric deviation models of interest, we provide the DE equations for three decoders: BP, Gallager-B, and MS. The Gallager-B and MS have not been considered in~\cite{wang2005density}, and~\cite{Huang2014Gallager} only proposed approximated DE equations for the Gallager-B decoder. 
 \item We propose to introduce asymmetric parameters onto Gallager-B and MS decoders, while in~\cite{Huang2014Gallager}, only symmetric Gallager-B decoders were considered under asymmetric deviation models. We then show how to optimize these asymmetric parameters so as to compensate for asymmetric deviations introduced in the decoder.
\end{enumerate}
}

The outline of the paper is as follows. Section~\ref{sec:ldpc_def} gives our assumptions and notations for LDPC decoders under asymmetric deviations. Section~\ref{sec:density_evolution_definition} introduces the noisy DE analysis under asymmetric deviations. Section~\ref{sec:density_evolution_equations} provides the noisy DE equations for the three considered decoders. Section~\ref{sec:optim} describes asymmetric decoder parameters. To finish, Section~\ref{sec:simu} provides simulation results.

\section{LDPC decoders}\label{sec:ldpc_def}
In this section, we first describe standard LDPC decoders without deviations. We then introduce the generic asymmetric deviation model we consider in this paper.
\textcolor{black}{Finally, we discuss practical decoder implementations to convey that the deviation models and the assumptions it contains are realistic.}

\subsection{Standard LDPC decoder without deviations}
We denote by $\Xseq$ a binary codeword of length $n$. The codeword $\Xseq$ is transmitted over a noisy \gls{iid} channel $\Pr(y|x)$ which outputs a vector $\Yseq$. For simplicity, we assume that the considered channel $\Pr(y|x)$ satisfies the symmetry conditions of~\cite{richardson2001capacity} which can be stated as $\Pr(y|x=1) = \Pr(-y|x=0) $.
We denote by \textcolor{black}{$\mathbf{H}$} the parity-check matrix of LDPC code, of size $m\times n$ and rate $R=m/n$.
\textcolor{black}{In this paper, for simplicity, most of the DE analysis is described for regular LDPC codes with VN degree $d_v$ and CN degree $d_c$, but we also show how to extend the DE analysis to irregular LDPC codes}. 

\textcolor{black}{We consider an LDPC decoder which performs $L$ iterations and which satisfies the extrinsic principle. The extrinsic principle means that when computing a message from VN $v$ to CN $c$, we use all the messages incoming to VN $v$, except the one coming from CN $c$.} In this decoder, the initial message incoming to a given VN is denoted $\minit$.  At iteration $\ell \in \{1,\cdots,L\}$, input CN messages \textcolor{black}{for a given CN} are denoted $(\mvtoc_{1}^{(\ell-1)}, \cdots, \mvtoc_{d_c-1}^{(\ell-1)}) $ and output CN message is denoted $\mctov_{d_c}^{(\ell)} $. The CN mapping is denoted $\Phic$ with
\begin{equation}\label{eq:CN_mapping}
 \mctov_{d_c}^{(\ell)} = \Phic(\mvtoc_{1}^{(\ell-1)}, \cdots, \mvtoc_{d_c-1}^{(\ell-1)}) .
\end{equation}
In the same way, input VN messages \textcolor{black}{for a given VN} are denoted $(\mctov_{1}^{(\ell)}, \cdots, \mctov_{d_v-1}^{(\ell)}) $ and the output VN message is denoted $\mvtoc_{d_v}^{(\ell)} $. 
The VN mapping is denoted $\Phiv$ with 
 \begin{equation}\label{eq:VN_mapping}
  \mvtoc_{d_v}^{(\ell)} = \Phiv(\minit, \mctov_{1}^{(\ell)}, \cdots, \mctov_{d_v-1}^{(\ell)}) .
 \end{equation}
 We also consider an \emph{a posteriori} probability (APP) mapping $\Phia $. In the decoder, the decision on the bit values is taken as  
 $$
\hat{x} = \left\{
    \begin{array}{ll}
        0 & \mbox{if } \Phia(\minit, \mctov_{1}^{(\ell)}, \cdots, \mctov_{d_v}^{(\ell)})>0 \\
        1 & \mbox{if } \Phia(\minit, \mctov_{1}^{(\ell)}, \cdots, \mctov_{d_v}^{(\ell)})<0
    \end{array}
\right.
 $$
 In addition, if $\Phia(\minit, \mctov_{1}^{(\ell)}, \cdots, \mctov_{d_v}^{(\ell)})=0$, then $\hat{x}$ is sampled uniformly at random.  
 The decoder stops after $L$ iterations or if the stopping condition \textcolor{black}{$\mathbf{H} \XseqE = 0$} is satisfied. 
Although it is often used in LDPC decoder implementations, the stopping condition is not taken into account in the DE analysis. 
 
 The above mappings $\Phiv$, $\Phic$, $\Phia$, give a generic description of a noiseless LDPC decoder which satisfies the extrinsic principle. We choose this description because in this paper, we consider several different decoders. In order to study a specific decoder such as Gallager~B or MS, it suffices to replace the mappings  $\Phiv$, $\Phic$, $\Phia$, by the ones of the considered decoder.

\subsection{Asymmetric deviation model}\label{sec:error_model}
 We now describe our main assumptions for the asymmetric deviation models we consider in this paper.
In the following, we denote by $\mvtocN_{i}^{(\ell)}$ ($i\in \{1,\cdots, d_{c-1}\}$), and by $\mctovN_{j}^{(\ell)}$ ($j\in \{1,\cdots, d_{v-1}\}$), the noisy versions of the messages $\mvtoc_{i}^{(\ell)}$ and $\mctov_{j}^{(\ell)}$ from VN to CN and from CV to VN, respectively. 
\textcolor{black}{Based on this notation, we first assume that hardware deviations are memoryless in the sense that deviations introduced at iteration $\ell$ are statistically independent of deviations introduced at previous iterations. This can be expressed as
\begin{align}\label{eq:memoryless_assumption1}
\Pr(\mvtocN_{i}^{(\ell)}|\mvtoc_{i}^{(\ell)}, \mvtocN_{i}^{(\ell-1)},\cdots \mvtocN_{i}^{(1)} ) & = \Pr(\mvtocN_{i}^{(\ell)}|\mvtoc_{i}^{(\ell)}), \\ \label{eq:memoryless_assumption2}
\Pr(\mctovN_{j}^{(\ell)}|\mctov_{j}^{(\ell)}, \mctovN_{j}^{(\ell-1)},\cdots \mctovN_{j}^{(1)} ) & = \Pr(\mctovN_{j}^{(\ell)}|\mctov_{j}^{(\ell)})
\end{align}
}
We also assume that deviations are applied on the noiseless mapping outputs $\mvtoc_{d_v}^{(\ell)}$ and $\mctov_{d_c}^{(\ell)}$. 
More formally, this corresponds to assuming that 
\begin{align}\label{eq:input_out1}
\Pr(\mvtocN_{d_v}^{(\ell)}|\mvtoc_{d_v}^{(\ell)}, \minit, \mctov_{1}^{(\ell)}, \cdots, \mctov_{d_v-1}^{(\ell)} ) & = \Pr(\mvtocN_{d_v}^{(\ell)}|\mvtoc_{d_v}^{(\ell)}), \\
\label{eq:input_out2}
\Pr(\mctovN_{d_c}^{(\ell)}|\mctov_{d_c}^{(\ell)}, \mvtoc_{1}^{(\ell)}, \cdots, \mvtoc_{d_c-1}^{(\ell)}) & = \Pr(\mctovN_{d_c}^{(\ell)}|\mctov_{d_c}^{(\ell)}).
\end{align}
Expressions~\eqref{eq:input_out1} and~\eqref{eq:input_out2} mean that, given the knowledge of the noiseless output, the noisy output is statistically independent of the mapping inputs. 
As a result, the effect of deviations on the decoder is entirely represented by the \textcolor{black}{conditional probability density function} $\Pr(\mvtocN_{d_v}^{(\ell)}|\mvtoc_{d_v}^{(\ell)})$ and $\Pr(\mctovN_{d_c}^{(\ell)}|\mctov_{d_c}^{(\ell)})$. 
Note that the \textcolor{black}{probability density functions} $\Pr(\mvtocN_{d_v}^{(\ell)}|\mvtoc_{d_v}^{(\ell)})$ and $\Pr(\mctovN_{d_c}^{(\ell)}|\mctov_{d_c}^{(\ell)})$ may vary from iteration to iteration.
The above two assumptions are performed in most works on LDPC decoders with deviations, see~\cite{varshney2011performance,al2014fault,sundararajan2014noisy,ngassa2015density,balatsoukas2014density,leduc2018modeling,Dupraz15Com}, \textcolor{black}{and Section~\ref{sec:implem} provides additional justification based on hardware implementation considerations}.

In addition, most existing works on faulty LDPC decoders~\cite{varshney2011performance,al2014fault,sundararajan2014noisy,ngassa2015density,balatsoukas2014density,Dupraz15Com} assume symmetric error models for which 
\begin{equation}\label{eq:sym_cdt1}
\Pr(\mvtocN_{d_v}^{(\ell)}|\mvtoc_{d_v}^{(\ell)}) = \Pr(-\mvtocN_{d_v}^{(\ell)}|-\mvtoc_{d_v}^{(\ell)}) \text{ and }  \Pr(\mctovN_{d_c}^{(\ell)}|\mctov_{d_c}^{(\ell)}) = \Pr(-\mctovN_{d_c}^{(\ell)}|-\mctov_{d_c}^{(\ell)}) . 
\end{equation}
A different symmetry condition is provided in~\cite{leduc2018modeling}. Assuming that the channel noise is represented by a vector $\mathbf{z}^n$ such that $y_i=x_i z_i$ for all $i \in \{1,\cdots,n\}$, the symmetry condition of~\cite{leduc2018modeling} is given by
\begin{equation}\label{eq:sym_cdt2}
 \Pr(\mvtocN_{d_v}^{(\ell)}|\mathbf{y}^n ) = \Pr(x \mvtocN_{d_v}^{(\ell)}|\mathbf{z}^n ) \text{ and } \Pr(\mctovN_{d_c}^{(\ell)}| \mathbf{y}^n ) = \Pr(-x \mctovN_{d_c}^{(\ell)}|\mathbf{z}^n),
\end{equation}
where $x\in\{-1,1\}$ is the codeword bit value associated to the VN that sends $\mvtocN_{d_v}^{(\ell)}$ or receives $\mctovN_{d_c}^{(\ell)}$. The two models described by~\eqref{eq:sym_cdt1} and~\eqref{eq:sym_cdt2} give that the decoder error probability is independent of the transmitted codeword. This allows to consider that the all-zero codeword was transmitted in both the theoretical analysis and in the Monte Carlo simulations of the decoder. 

Here, unlike in~\cite{varshney2011performance,al2014fault,sundararajan2014noisy,ngassa2015density,balatsoukas2014density,leduc2018modeling,Dupraz15Com}, we do not perform any assumption on the symmetry of the \textcolor{black}{probability density functions} of $\mvtocN_{d_v}^{(\ell)}$ and $\mctovN_{d_c}^{(\ell)}$. As a result, we cannot rely on the all-zero codeword assumption. 
In \textcolor{black}{Section~\ref{sec:density_evolution_definition}}, we propose a noisy DE analysis that allows to evaluate the performance of LDPC decoders under asymmetric deviation models without considering the all-zero codeword assumption.

{\color{black}
\subsection{Faulty Decoder Implementations}\label{sec:implem}

To better place the proposed deviation model in context, let us briefly review some relevant aspects of a decoder implementation.

\subsubsection{Asymmetric deviations}
The need to study asymmetric deviations, that is deviations for which the probability of confusing a logic 0 for a 1 is different from the probability of confusing a logic 1 for a 0, arises from the fact that digital systems are commonly engineered from physical mechanisms that are asymmetric in nature.
For instance, the vast majority of digital systems are currently built out of CMOS circuits. In CMOS, a logic gate is composed of PMOS devices used for pulling the output up to logic 1, and of NMOS devices used for pulling the output down to logic 0, but PMOS and NMOS devices have different physical properties and must be designed differently.
The situation is common in data storage as well. An often used mechanism called \emph{dynamic} memory consists in storing an electric charge to represent a logic 1, but removing the charge to represent a logic 0. This mechanism is also asymmetric in nature since a charge is more likely to leak than to spontaneously appear. For instance, embedded dynamic random-access memories (eDRAMs) are seen as a promising approach to increase storage densities of on-chip memories, but the retention time of a bit stored in an eDRAM cell can be very different depending on whether a $0$ or a $1$ is stored \cite[Fig.~4]{Bravo:2019}.
Even \emph{static} memories sometimes use \emph{single-ended} (and thus asymmetric) designs to balance the various conflicting requirements of the design problem~\cite{verma:2008}.
In all these cases, circuit designers usually aim to restore symmetry by careful design, but this comes at a cost. For instance, \cite{nabavi:2016} shows that optimizing the sum of rise and fall times in a CMOS logic gate operated in the energy-efficient \emph{subthreshold} regime is achieved with unequal rise and fall times. As a result, allowing timing violations in such a circuit would result in asymmetric deviations.
Perhaps more importantly, as fabrication variations become increasingly problematic when designing energy-efficient circuits, symmetry is one of the engineered features of the system that will tend to break down.

\subsubsection{Designing a faulty implementation}
%

In order to use the proposed analysis to design a faulty decoder implementation, it is necessary to ensure that the assumptions of Section~\ref{sec:error_model} are verified.
At first, the memoryless assumptions~\eqref{eq:memoryless_assumption1}, ~\eqref{eq:memoryless_assumption2},
might seem to be violated by the fact that deviations can be due to fabrication variations, which are permanent. But an important distinction must be made between physical circuits and decoder messages. Indeed, a different computing circuit can be used to perform the computations associated with a particular Tanner graph node at different iterations of the algorithm. Similarly, a message associated with a particular edge of the Tanner graph can be stored at different memory locations at different times.

Assumptions~\eqref{eq:memoryless_assumption1}, ~\eqref{eq:memoryless_assumption2}
remain adequate in a decoder implementation affected by permanent faults if computations or message storage operations are realized by drawing at random a computation unit or a memory unit, as the case may be, from a large pool of such units. This behavior can be approximated in practice by enforcing in the implementation a \emph{process diversity} policy, that is by ensuring through a combination of system architecture and control assignments that related parts of a computation are performed by different physical units.
Fortunately, process diversity can often be provided easily in practical LDPC decoder architectures. For instance, in a parallel decoder architecture targeted at the widely used quasi-cyclic LDPC code family~\cite{nadal:2020a}, messages are routed between two pools of processing units by applying rotations to vectors of messages.
The rotation units could serve the additional purpose of assigning a given node to different computing or memory circuits in different iterations so as to approximate memoryless deviations.

On the other hand, depending on the type of deviation that is studied, there is no guarantee that \eqref{eq:input_out1} and \eqref{eq:input_out2} are verified. When they are not, they should be seen as a necessary approximation to make the DE tractable, with the redeeming factor that this approximation can be made as precise as desired since a different deviation distribution can be used for every DE iteration.

} 

\section{Noisy Density Evolution with Asymmetric Deviation Models}\label{sec:density_evolution_definition}
In this section, we propose a noisy DE analysis of LDPC decoders under asymmetric deviation models. In particular, we introduce the mathematical formalism that allows to perform DE without the all-zero codeword assumption. 
In our analysis, we consider the cycle-free assumption which is usually used to derive DE equations~\cite{richardson2001capacity,richardson2001design}. This assumption implies that the messages incoming to a VN or a CN are statistically independent. 

\subsection{Error probability evaluation}
We now describe the DE analysis that allows to evaluate the message error probability of the decoder at successive iterations. 
We denote by $\Pv_x(\mvtoc)$ the \textcolor{black}{probability density function} of a noiseless message $\mvtoc$ at a VN output, conditioned on the codeword bit $x\in\{0,1\}$ at this VN. 
We also denote $\Pc_x(\mctov)$ the \textcolor{black}{probability density function} of a noiseless message $\mctov$ at a CN output conditioned on the codeword bit $x$ at the VN that receives the message $\mctov$. 
The notations $\PvN_x(\mvtocN)$ and $\PcN_x(\mctovN)$ hold for the  \textcolor{black}{probability density functions} of noisy messages $\mvtocN$ and $\mctovN$.
\textcolor{black}{Note that we use the term probability density function, although in the paper we consider both continuous-valued and discrete-valued decoder messages. In the latter case, the density function can be defined from the counting measure and reduces to a probability mass function. }
Here, unlike in~\cite{Huang2014Gallager}, we consider conditioning on the bit value $x$ rather than on the codeword weight $w$. This is because~\cite{wang2005density} shows that the two approaches are equivalent, while the terms $\Pv_0(\mvtoc)$, $\Pv_1(\mvtoc)$, etc. are simpler to evaluate when conditioned on $x$ rather than on $w$. 

DE consists of calculating the \textcolor{black}{probability density functions} $\Pv_0(\mvtoc)$, \textcolor{black}{$\Pv_1(\mvtoc)$}, etc., at successive iterations $\ell$. 
The commonly considered symmetry assumptions restated in Section~\ref{sec:error_model} lead to the equalities $\Pv_0(\mvtoc) = 1 - \Pv_1(\mvtoc)$, $\Pc_0(\mctov) = 1 - \Pc_1(\mctov) $, and the same holds for $\PvN_x(\mvtocN)$ and $\PcN_x(\mctovN)$.
In standard DE, these equalities allow to consider the all-zero codeword assumption, and as a result, only the terms $\Pv_0(\mvtoc)$, $\Pc_0(\mctov)$ (for noiseless messages), $\PvN_0(\mvtocN)$, $\PcN_0(\mctovN)$ (for noisy messages), are calculated. On the contrary, here, such equalities do not hold and it is required to calculate all the terms $\Pv_0(\mvtoc), \Pv_1(\mvtoc)$, $\Pc_0(\mctov)$, $\Pc_1(\mctov)$, and their noisy counterparts.

When the all-zero codeword assumption is removed, we can express the message error probability $\Pe$ of the decoder as follows. 
First denote 
\begin{equation}\label{eq:expP}
\PvN(\mvtocN) = \frac{1}{2} \PvN_0(\mvtocN) + \frac{1}{2} \PvN_1(-\mvtocN).
\end{equation} 
\textcolor{black}{In this expression, the  probability density functions $\PvN_0(\mvtocN)$ and $\PvN_1(-\mvtocN)$ are weighted with a coefficient $1/2$ because we assume that the transmitted codeword bits are equiprobable, which is the capacity-achieving distribution when the channel is symmetric}. We then have that the message error probability $\Pe$ in the noisy decoder at iteration $\ell$ can be expressed as
 \begin{equation}\label{eq:error_prob}
  \Pe = \int_{-\infty}^{0-} \PvN(\mvtocN) d\mvtocN + \frac{1}{2}\textcolor{black}{\PvN(0)},
 \end{equation}
 \textcolor{black}{where $\PvN(0)$ is the probability that a noisy message $\mvtocN$ takes value $0$. In the above expression, for continuous-valued messages, the second term is often equal to $0$, unless there is some mass point at $0$ (this would be the case, for instance, for the Self-Corrected Min-Sum decoder introduced in~\cite{savin2008self}). For discrete-valued messages,~\eqref{eq:error_prob} reduces to $\Pe = \sum_{\mvtocN < 0} \PvN(\mvtocN) +  \frac{1}{2}\textcolor{black}{\PvN(0)} $. Also note that $\Pe$ depends on the probability density function   $\PvN(\mvtocN)$ whose expression itself depends on the considered decoder. The expression of $\PvN(\mvtocN)$ will be specified later in the paper for different decoders.}
 
An expression similar to~\eqref{eq:error_prob} was considered in~\cite{wang2005density,dupraz2014density} in the case of an asymmetric communication channel, and asymmetric VN and CN mappings, but without deviations in the decoder. 
 In the expression of $\Pe$ in~\eqref{eq:error_prob}, the second term comes from the case where the APP decision gives $\Phia(\minit, \mctov_{1}^{(\ell)}, \cdots, \mctov_{d_v}^{(\ell)}) = 0$.
In addition, considering $\PvN_1(-\mvtocN)$ rather than $\PvN_1(\mvtocN)$ in the definition of $\PvN(\mvtocN)$ in~\eqref{eq:expP} allows to have one single integral for the first term of $\Pe$.

As a result, in order to evaluate the message error probability $\Pe$, we need to express the  \textcolor{black}{probability density functions} $\PvN_0(\mvtocN)$ and $\PvN_1(\mvtocN)$ for the two possible codeword bit values $x=0$ and $x=1$.
The expressions of these  \textcolor{black}{probability density functions} depend on the considered LDPC decoder. In Section~\ref{sec:density_evolution_equations}, we give their expressions for three decoders: BP, Gallager~B, and MS. 

\subsection{Threshold definition}\label{sec:th_def}
The message error probability $\Pe$ defined in~\eqref{eq:error_prob} is calculated under particular channel and deviation models. The code ensemble threshold~\cite{richardson2001capacity} permits to evaluate the average decoder performance for the ensemble of all codes with regular degrees $d_v$ and $d_c$. 
For decoders without deviations, the threshold is evaluated as the worst channel parameter that allows to have a vanishing error probability $p_e^{(\ell)}$ when the codeword length tends to infinity~\cite{richardson2001capacity}. 

For decoders with deviations, the standard threshold definition does not apply. Indeed, deviations usually prevent the decoder from reaching a zero error probability.  Therefore, several alternative threshold definitions were introduced for decoders with deviations~\cite{varshney2011performance,Dupraz15Com}.
In this paper, we consider the definition of~\cite{varshney2011performance} which sets up a parameter $\epsilon$ and defines the threshold as the worst channel parameter for which $\Pe < \epsilon$. Although introduced in the case of symmetric deviation models, this definition still applies to the case of asymmetric deviations. 

\subsection{Finite-length evaluation of the message error probability}\label{sec:fl_de}
The error probability $\Pe$ defined in~\eqref{eq:error_prob} is evaluated under asymptotic conditions, that is assuming that the codeword length $n$ goes to infinity. 
Alternatively,~\cite{leduc2016finite} describes a method for evaluating the error probability at finite length as follows. 

We first describe the method of~\cite{leduc2016finite} for a binary symmetric channel (BSC) with parameter $p_0$. 
We denote by $\PeN$ the message error probability for a codeword length $n$. For a given value $p_0$, the error probability $\PeN(p_0)$ can be evaluated as
\begin{equation}\label{eq:pe_fl}
 \PeN(p_0) = \int_{0}^{1/2} \Pe(z) \, \Phi_{\mathcal{N}}\! \left(z; p_0, \frac{p_0(1-p_0)}{n}\right) dz \, .
\end{equation}
In  this expression, $\Pe(z)$ is the asymptotic error probability given in~\eqref{eq:error_prob}, \textcolor{black}{whose expression will be specified in Section~\ref{sec:density_evolution_equations}, depending on the considered decoder. } In addition, $\Phi_\mathcal{N}$ gives the probability density function of a Gaussian random variable with mean $p_0$ and variance $\frac{p_0(1-p_0)}{n}$. 
This expression is obtained by considering that the observed BSC parameter can be modeled as a Gaussian random variable~\cite[Chapter 8]{friedman2001elements}.  We can apply the same method in order to obtain the error probability $\PeN(\sigma^2)$ for an Additive White Gaussian Noise (AWGN) channel of variance $\sigma^2$. For this, it suffices to apply~\eqref{eq:pe_fl} with $p_0 = \frac{1}{2} - \frac{1}{2} \text{erf} \left( \frac{1}{\sqrt{2\sigma^2}} \right)$, where $\text{erf}$ is the error function of the Gaussian distribution. In this case, in the integral in~\eqref{eq:pe_fl}, the asymptotic error probability $\Pe(z)$ is evaluated for a channel variance $v^2$ that depends on $z$ and is given by $v^2 = \frac{1}{ 2\left( \text{erf}^{-1} (1-2z) \right)^2  }$.

This method allows to take into account the channel variations at finite-length. However, it still assumes that the code is cycle-free, and hence does not evaluate the effect of cycles onto the code performance. Since cycles are known to degrade the code performance at short length~\cite{hu2005regular}, the method presented in this section is well-suited to evaluate the finite-length performance for moderate to long lengths.
We will use this method in our experiments in order to verify the accuracy of the DE equations that we now introduce for three decoders: BP decoder, Gallager-B decoder, and quantized  MS decoder.

\section{Noisy Density Evolution Equations}\label{sec:density_evolution_equations}
The DE methodology described in the previous section is generic and may be applied to any type of LDPC decoder and deviation model that satisfies the assumptions of Section~\ref{sec:ldpc_def}. 
In this section, we study three of the most common LDPC decoders: BP decoder, Gallager~B decoder, and quantized MS decoder.  
For each considered decoder, we give examples of asymmetric deviation models and derive the corresponding noisy DE equations. 
\textcolor{black}{In this part, we derive the DE equations for regular $(d_v,d_c)$ LDPC codes for the three decoders. We then show how to extend the analysis to irregular LDPC codes. }

\subsection{BP decoder}\label{sec:bp_dec}
We first consider the standard BP decoder with infinite precision on the messages. 
This first derivation of DE equations under asymmetric deviation models is mostly of theoretical interest since this decoder cannot be directly implemented in hardware.
In the BP decoder, the messages are initialized with the channel log-likelihood ratios (LLRs) as $\minit = \log \frac{\Pr(X=0|y)}{\Pr(X=1|y)} . $
The VN and CN mappings~\eqref{eq:CN_mapping} and~\eqref{eq:VN_mapping} are as follows:
 \begin{align}
 \mvtoc^{(\ell)}_{d_v} =& \, \minit + \sum_{j=1}^{d_v-1} \mctov^{(\ell-1)}_{j} \label{eq:BP_dec_vn} \\
  \mctov^{(\ell)} =& \, g^{-1}\! \left( \sum_{i=1}^{d_c-1} g\!\left(  \mvtoc^{(\ell)}_i \right) \right) . \label{eq:BP_dec}  
 \end{align}
 \textcolor{black}{where, according to~\cite{richardson2001design}, the function $g:\mathbb{R} \rightarrow \mathbb{R}^2$ is defined as $g(x) = (\text{sgn}(x), -\log \tanh |\frac{x}{2}|)$, and $\text{sgn}(\cdot)$ is the sign function.}

For the BP decoder, we can consider for instance the additive deviation model described in~\cite{tarighati2015design,varshney2011performance}.  In this model, noisy VN output messages $\mvtocN^{(\ell)}_{d_v}$ are in the form 
\begin{equation}\label{eq:BP_VN_noise}
\mvtocN^{(\ell)}_{d_v}  = \mvtoc^{(\ell)}_{d_v} + b^{(\ell)},
\end{equation} where $b^{(\ell)}$ is a continuous random variable that represents the noise, \textcolor{black}{and the noise on CN output messages $\mctov^{(\ell)}_{d_c}$ is also additive, and as such can be combined with additive noise on VN output messages without loss of generality since the VN mapping~\eqref{eq:BP_dec_vn} is a sum.}
The random variable $b^{(\ell)}$ is assumed independent of the noiseless message value $\mvtoc^{(\ell)}_{d_v}$. 
The probability density function $B$ of $b^{(\ell)}$ does not depend on $\ell$. For instance,~\cite{tarighati2015design} considers a zero-mean Gaussian distribution for $b^{(\ell)}$. However, here, unlike in~\cite{tarighati2015design,varshney2011performance}, no symmetry assumption is placed on $B$, which means that we might have $B(\mvtocN) \neq B(-\mvtocN)$ (for instance, a $\chi^2$-distribution).

We now give the DE equations for the BP decoder with the above additive deviation model, \textcolor{black}{which generalize the asymmetric DE equations from~\cite{wang2005density} to handle deviations}. First, the \textcolor{black}{probability density functions} of the initial messages $\minit$ depends on the channel model. For instance, for an AWGN channel, $\minit=2y/\sigma^2$, where $y$ is the channel output and $\sigma^2$ is the channel variance. In this case, the probability density function $\Pinit_x$ of $\minit$ is a Gaussian distribution with mean $2(1-2x)/\sigma^2 $, where $x\in \{0,1\}$, and variance $4/\sigma^2$.
Then, the probability density function of the noisy VN output messages $\mvtocN$ can be expressed as
\begin{equation}\label{eq:probaVN_BP}
 \PvN_x(\mvtocN) = \Pinit_x \otimes \left( Q_x^{(\ell-1)} \right)^{\otimes(d_v-1)}  \otimes B(\mvtocN) ,
\end{equation}
where $\otimes$ represents the convolution product, and $(.)^{\otimes d}$ represents the power convolution operator. By convention, $(.)^{\otimes 0} = 1$.
Next, the probability density function of the noiseless CN output messages $\mctov$ is given by
\begin{align}\label{eq:probaCN_BP}
 \Pc_0(\mctov) & = \left( \frac{1}{2} \right)^{d_c-2} \sum_{v=0, \text{ even }}^{d_c-1} \binom{d_c-1}{v} \Gamma^{-1} \left( \Gamma\left( \PvN_0 \right)^{\otimes(d_c-1-v)} \otimes \Gamma\left( \PvN_1 \right)^{\otimes v} \right) \\
 \Pc_1(\mctov) & = \left( \frac{1}{2} \right)^{d_c-2} \sum_{v=1, \text{ odd }}^{d_c-1} \binom{d_c-1}{v} \Gamma^{-1} \left( \Gamma\left( \PvN_0 \right)^{\otimes(d_c-1-v)} \otimes \Gamma\left( \PvN_1 \right)^{\otimes v} \right)
\end{align}
where $\Gamma$ is the density transform operator of the function $g$, see~\cite{chen2009equivalence}. 
The VN output messages probability density function~\eqref{eq:probaVN_BP} is obtained from the fact that the VN computation~\eqref{eq:BP_dec_vn} is a sum of independent messages.
For the derivation of the CN output message probability density function~\eqref{eq:probaCN_BP}, see Appendix~\ref{app:proofCN}.
\textcolor{black}{Finally, note that the density transform operator $\Gamma$ and its inverse $\Gamma^{-1}$ do not have known analytical expressions. Therefore, for the BP decoder, DE is usually evaluated either from message quantization~\cite{richardson2001design} or from Monte-Carlo simulations~\cite{gorgoglione2010optimized}. }


\subsection{Gallager~B decoder}\label{sec:gb_dec}
We now consider the Gallager~B decoder with a BSC of parameter $p_0$ and under an asymmetric deviation model. The Gallager~B decoder works with hard-decision messages $\mvtoc,\mctov \in \{0,1\} $. 
In this decoder, the initial message $\minit$ is equal to the channel output, that is $\minit = y$, with $y \in \{0,1\}$. 
The VN mapping is given by
\begin{equation}\label{eq:VN_GB}
\mvtoc_{d_v}^{(\ell)} = \left\{
    \begin{array}{ll}
        \minit \oplus 1 & \mbox{if } \left| \{ d: \mctov_d^{(\ell-1)} = \minit \oplus 1 \} \right| \geq \thgb \\
        \minit & \mbox{otherwise.}
    \end{array}
\right.
\end{equation}
where $\thgb$ is a decoder parameter that depends on the iteration number $\ell$ and can be optimized with DE. 
The CN mapping is given by
 \begin{equation}\label{eq:CN_GB}
  \mctov_{d_c}^{(\ell)} = \bigoplus_{d=1}^{d_c-1} \mvtoc_d^{(\ell)} ,
 \end{equation}
 where $\bigoplus$ denotes the XOR sum of the $\mvtoc_d^{(\ell)}$.

For the Gallager~B decoder, we can consider the same deviation model as in~\cite{Huang2014Gallager} and apply it on the output CN messages $\mvtoc_d$.
This model is described by two parameters $\epp$ and $\epn$ such that $\epp = \Pr(\mctovN_d=1|\mctov_d=0)$ and $\epn = \Pr(\mctovN_d=0|\mctov_d=1)$, where we do not assume that $\epp = \epn$. 
\textcolor{black}{Note that since the CN message computation~\eqref{eq:CN_GB} is a XOR sum, it is easy to show that deviations applied onto VN messages could be combined with deviations applied at the output CN messages. Therefore, the above deviation model which only considers some noise on output CN messages can be considered without loss of generality. }

With this model, the DE equations of the Gallager~B decoder are as follows. The \textcolor{black}{probability density function} of initial messages $\minit$ is $\Pinit_0(1) = p_0$ and $\Pinit_1(1) = 1-p_0$.
Then, the \textcolor{black}{probability density functions} of the noiseless CN output messages $\mctov_{d_c} \in \{0,1\}$ are given by 
\begin{align}
 \Pc_0(1) & =  \left( \frac{1}{2} \right)^{d_c-1} \sum_{v=0, \text{ even }}^{d_c-1} \binom{d_c-1}{v} \left( 1 - \left(1-2\Pv_0(1)\right)^{d_c-1-v}\left(1-2\Pv_1(1)\right)^v\right) \\ \label{eq:CN_pmf_GB}
 \Pc_1(1) & = \left( \frac{1}{2} \right)^{d_c-1} \sum_{v=1, \text{ odd }}^{d_c-1} \binom{d_c-1}{v} \left( 1 - \left(1-2\Pv_0(1)\right)^{d_c-1-v}\left(1-2\Pv_1(1)\right)^v\right)
\end{align}
The derivation of these expressions can be done by following the same steps as for the BP decoder, see Appendix~\ref{app:proofCN}.
The \textcolor{black}{probability density functions} of the noisy CN output messages $\mctovN_{d_v} \in \{0,1\}$ can be expressed as
\begin{equation}\label{eq:GB_noise}
 \PcN_x(0)  = \epn \Pc_x(1) + (1-\epp) \Pc_x(0) .
\end{equation}
From the condition $\epp \neq \epn$, we can show that $\PcN_0(0) \neq 1- \PcN_1(0)$ in general.
The \textcolor{black}{probability density functions} of the noiseless VN output messages $\mvtoc_d \in \{0,1\}$ are given by
\begin{align}\notag
 \Pv_x(0) = & \Pinit_x(0)\sum_{v=\thgb}^{d_v-1} \binom{d_v-1}{v}\left(1-\PcN_x(1)\right)^{d_v-1-v}\PcN_x(1)^v  \\ \label{eq:VN_pmf_GB}
 & + \Pinit_x(1) \left(1-\sum_{v=\thgb}^{d_v-1} \binom{d_v-1}{v}\left(1-\PcN_x(1)\right)^{v}\PcN_x(1)^{d_v-1-v}\right) ,
\end{align}
and  $\Pv_x(1) = 1 -  \Pv_x(0)$. 
From these expressions, one can determine the code ensemble threshold and optimize the decoder parameter $\thgb$.

Note that here, we considered messages in a binary alphabet $\{0,1\}$. Therefore, the error probability calculation~\eqref{eq:error_prob} cannot be applied. However, we can easily evaluate the Gallager~B decoder error probability as
\begin{equation}\label{eq:error_prob_GB}
 \Pe = \frac{1}{2} \Pv_0(1) + \frac{1}{2} \Pv_1(0) .
\end{equation}

\subsection{Quantized Offset MS decoder}\label{sec:dec_ms}
In this section, we consider a quantized offset MS decoder, which is often considered in practical hardware implementations~\cite{leduc2018modeling}.
For message quantization in the decoder, we consider a uniform quantizer on $q$ bits, which gives $2^q-1$ quantization levels. The length of the quantization intervals is $\mu$ and the message quantization alphabet is $\M = \{ -2^{q-1}+1, \cdots, 0, \cdots,  2^{q-1}-1\}$. We denote by $\Q(.)$ the quantization function, with
\begin{equation}
  \Q(\mi) \left\{
    \begin{array}{ll}
         2^{q-1}-1  & \mbox{ if } \mi >  2^{q-1}-1  \\
        -2^{q-1}+1 & \mbox{ if } \mi <   -2^{q-1}+1 \\
       \mu \left\lfloor \frac{m}{\mu} + \frac{1}{2} \right\rfloor & \mbox{ otherwise. }
    \end{array}
\right.
\end{equation}

For an AWGN channel with variance $\sigma^2$, the messages are initialized as $\minit = \Q(2\alp y/\sigma^2)$, where $y$ is the channel output and $\alp$ is a scaling factor.
The VN mapping is then given by
 \begin{equation}\label{eq:VNqMS}
   \mvtoc^{(\ell)}_{d_v} = \Q\left( \minit + \sum_{d=1}^{d_v-1} \mctov^{(\ell-1)}_{d} \right) ,
 \end{equation}
where the quantization function $\Q$ only serves to perform message saturation. 
 The CN mapping is given by
\begin{equation}\label{eq:CN_MS}
 \mctov_{d_c}^{(\ell)} =    \left(\prod_{d=1}^{d_c-1} \text{sgn} \left(  \mvtoc^{(\ell)}_d \right) \right) \max\left( \min_{d} \left| \mvtoc^{(\ell)}_d  \right| - \off , 0 \right) 
\end{equation}
where $\text{sgn}(\cdot)$ is the sign function, and $\off$ is the decoder offset parameter.
 
Here, we consider a deviation model at the bit level for the quantized messages $ \mvtoc^{(\ell)}_{d_v}$.
We use $(b_1,\cdots, b_q)$ to represent the $q$ bit values of a symbol $\mi \in \M$. 
Then, as for the Gallager~B decoder described in Section~\ref{sec:gb_dec}, the deviation model is defined by the two parameters $\epp = \Pr(\tilde{b}_j=1|b_j=0)$ and $\epn = \Pr(\tilde{b}_j=0|b_j=1)$, where $\tilde{b}_j$ is the noisy version of $b_j$.  For simplicity, we assume that the noise parameters $\epp$ and $\epn$ are the same for all bits $b_j$.
In order to represent this deviation model for the quantized decoder, we can construct a probability transition matrix $\Pt$ of size $(2^q-1)\times(2^q-1)$, as initially proposed in~\cite{Dupraz15Com}. However,~\cite{Dupraz15Com} requires symmetry conditions for $\Pt$, which are unnecessary here.  
With the above deviation model, the components of the matrix $\Pt$ are such that $$\Pt_{i,k} = \Pr(\mvtocN^{(\ell)}_{d_v}=k | \mvtoc^{(\ell)}_{d_v}=i),$$ with $i,k \in \M$. 

For the defined decoder and deviation model, the DE equations are as follows. The initial message \textcolor{black}{probability density functions} $\Pinit_x$ can be evaluated by using the erf function, the error function of the Gaussian distribution, and depend on the quantization intervals. 
The probability density function of the noiseless VN output messages $\mvtoc_{d_v}^{(\ell)}$ is then given by 
\begin{equation}\label{eq:DE_VN_MS}
 \Pv_x(\mvtoc) = \sum_{\boldsymbol{\beta}: \Phiv(\boldsymbol{\beta}) = \mvtoc} \Pinit_x(\minit) \prod_{d=1}^{d_v-1} \Pc_x(\mctov_d),
\end{equation}
where $\boldsymbol{\beta} = (\minit,\mctov_1,\cdots,\mctov_{d_v-1})$.  
 Then, we use $\PvV_x = [ \Pv_x(-2^{q-1}+1),\cdots, \Pv_x(2^{q-1}-1)  ]$ to denote the probability density function in vectorial form of the noiseless VN output messages $\mvtoc_{d_v}^{(\ell)}$. With this notation, the probability density function in vectorial form $\PvNV_x$ of the noisy VN output messages $\mvtocN$ is given by 
 \begin{equation}\label{DE_VN_MS_Noisy}
 \PvNV_x = \Pt \PvV_x.
 \end{equation}
 Note that this expression only depends on the transition matrix $\Pt$ and could therefore be applied to other deviation models.
The probability density function of the noiseless CN output messages $\mctovN$ is given by
\begin{align}\label{eq:DE_CN_MS0}
 \Pc_0(\mctov) & = \left( \frac{1}{2} \right)^{d_c-2} \sum_{v=0, \text{ even }}^{d_c-1} \binom{d_c-1}{v} \sum_{ \boldsymbol{\alpha}: \Phic(\boldsymbol{\alpha}) = \mctov  } \prod_{d=1}^v \PvN_1(\mvtoc_d) \prod_{d=v+1}^{d_c-1} \PvN_0(\mvtoc_d)  \\\label{eq:DE_CN_MS1}
 \Pc_1(\mctov) & = \left( \frac{1}{2} \right)^{d_c-2} \sum_{v=1, \text{ odd }}^{d_c-1} \binom{d_c-1}{v}  \sum_{ \boldsymbol{\alpha}:  \Phic(\boldsymbol{\alpha}) = \mctov  } \prod_{d=1}^v \PvN_1(\mvtoc_d) \prod_{d=v+1}^{d_c-1}\PvN_0(\mvtoc_d),
\end{align}
where $\boldsymbol{\alpha} = (\mvtoc_1,\cdots,\mvtoc_{d_v-1})$. By convention, in the above expressions, if $v=0$, we set  $\prod_{d=1}^v \PvN_1(\mvtoc_d) = 1$, and if $v=d_c-1$, we set $\prod_{d=v+1}^{d_c-1} \PvN_0(\mvtoc_d) = 1$. The derivation of these expressions can be done by following the same steps as for the BP decoder, see Appendix~\ref{app:proofCN}.
From these expressions, one can determine the code ensemble threshold and optimize the decoder parameters $\alp$ and $\off$.
\textcolor{black}{Finally, the DE equations given in this section could be easily extended to the case where deviations appear at CN outputs, by considering a second transition matrix to be applied to a vector of CN output messages probabilities.}

\color{black}
\subsection{Extension to irregular LDPC codes}
We now describe how to extend the proposed DE analysis to the case of irregular LDPC codes. We consider irregular LDPC codes with edge-perspective VN degree distribution $\lambda(x)$ and with edge-perspective CN degree distribution $\rho(x)$~\cite{richardson2001design}. Following the notation of~\cite{richardson2001design}, the VN degree distribution can be expressed as $\lambda(x) = \sum_{i\geq 2}\lambda_i x^{i-1}$, where $\lambda_i$ is the fraction of edges emanating from a VN of degree $i$, and the CN degree distribution is given by $\rho(x) = \sum_{j\geq2} \rho_j x^{j-1}$, where $\rho_j$ is the fraction of edges emanating from a CN of degree $j$.

Then, for each of the three considered decoders, we denote by $\PvN_{x,i}(\tilde{\alpha})$ the probability density function of the noisy VN output messages, evaluated for $d_v=i$. For instance, for the BP decoder, $\PvN_{x,i}(\alpha)$ is calculated from~\eqref{eq:probaVN_BP} by setting $d_v=i$, but by still using $\Pc_x$. We also denote by $\Pc_{x,j}(\beta)$ the probability density function of the CN output messages, evaluated with $d_c=j$, from $\PvN_{x}(\alpha)$.
Then, at iteration $\ell$, the probability density functions of decoder messages have expressions
\begin{align}
 \PvN_{x}(\tilde{\alpha}) & = \sum_{i\geq2} \lambda_i \PvN_{x,i}(\alpha) \\
 \Pc_x(\beta) & = \sum_{j\geq2} \rho_j \Pc_{x,j}(\beta) .
\end{align}
In the simulation section, we mainly focus on regular LDPC codes, but also evaluate the performance of a few irregular LDPC codes from the proposed DE analysis. 

\color{black}

\section{Asymmetric Decoder Parameters}\label{sec:optim}
In this section, we want to optimize the decoder parameters under asymmetric error models. We focus on the two practical LDPC decoders that were introduced in Section~\ref{sec:density_evolution_equations}: the Gallager~B decoder and the offset MS decoder.
We propose variants of these two decoders, in which the parameter $\thgb$ of the Gallager~B decoder and the parameters $\alp, \off$ of the MS decoder are asymmetric in the sense that they now depend on the signs of the messages exchanged in the decoder. We show how to optimize these asymmetric parameters in order to improve the decoding performance under asymmetric error models.  

\subsection{Asymmetric parameters in the Gallager~B decoder}
The standard Gallager~B decoder described in Section~\ref{sec:gb_dec} depends on one parameter $\thgb$ which can vary from iteration to iteration. 
We now set two different parameters $\thgbZ$ and $\thgbO$ and re-define the Gallager~B VN update~\eqref{eq:VN_GB} as
\begin{equation}\label{eq:VN_GB_var}
\mvtoc_{d_v}^{(\ell)} = \left\{
    \begin{array}{ll}
         1 & \mbox{if } \minit = 0 \text{ and } \left| \{ d: \mctov_d^{(\ell)} = 1 \} \right| \geq \thgbZ, \\
         0 & \mbox{if } \minit = 1 \text{ and } \left| \{ d: \mctov_d^{(\ell)} = 0 \} \right| \geq \thgbO, \\
        \minit & \mbox{otherwise.}
    \end{array}
\right.
\end{equation}
In this equation, the parameter $\thgbZ$ is associated to channel output $\minit=0$ while the parameter $\thgbO$ is associated to channel output $\minit = 1$.
The Gallager~B CN update given in~\eqref{eq:CN_GB} does not change. 

With this new VN update, the VN message \textcolor{black}{probability density functions} $\Pv_x(0)$ in~\eqref{eq:VN_pmf_GB} now depend on both parameters $\thgbZ$ and $\thgbO$, and become
\begin{align}\notag
 \Pv_x(0) = & \Pinit_x(0)\sum_{v=b_0}^{d_v-1} \binom{d_v-1}{v}\left(1-\PcN_x(1)\right)^{d_v-1-v}\PcN_x(1)^v  \\
 & + \Pinit_x(1) \left(1-\sum_{v=b_1}^{d_v-1} \binom{d_v-1}{v}\left(1-\PcN_x(1)\right)^{v}\PcN_x(1)^{d_v-1-v}\right) .
\end{align}
The expressions of the other \textcolor{black}{probability density functions} $\PvN_x(0)$ and $\Pc_x(1)$ do not change compared to Section~\ref{sec:gb_dec}. 

We now propose a method to optimize the two parameters $\thgbZ$ and $\thgbO$. For this, we first extend the method of~\cite{richardson2001capacity,Huang2014Gallager} that only considers one single parameter $\thgbZ = \thgbO = \thgb$. 
Denote $\mathcal{H}_x\left(\thgbZ,\thgbO\right) = \Pv_x(0)$, where $\Pv_x(0)$ is evaluated with given parameters $\thgbZ$ and $\thgbO$. The objective is to find the smallest integers $\thgbZ$ and $\thgbO$ such that 
\begin{align}
 \mathcal{H}_0\left(\thgbZ+1,\thgbO+1 \right) & \geq \mathcal{H}_0\left(\thgbZ,\thgbO\right), \\
 \mathcal{H}_1\left(\thgbZ+1,\thgbO+1\right) & \leq \mathcal{H}_1\left(\thgbZ,\thgbO\right)  .
\end{align}
The first inequality comes from the fact that we want to maximize $\Pv_0(0)$ and the second one comes from the fact that we want to minimize $\Pv_1(0)$. 
These two inequalities can be restated as
\begin{align}\label{eq:ineq1_param}
 \frac{\Pinit_0(1)}{\Pinit_0(0)} & \geq \frac{\binom{d_v-1}{\thgbZ}}{\binom{d_v-1}{\thgbO}} \left(\frac{1-\PcN_0(1)}{\PcN_0(1)} \right)^{d_v-1-\thgbZ - \thgbO}, \\
\label{eq:ineq2_param}
     \frac{\Pinit_1(1)}{\Pinit_1(0)} & \leq \frac{\binom{d_v-1}{\thgbZ}}{\binom{d_v-1}{\thgbO}} \left(\frac{1-\PcN_1(1)}{\PcN_1(1)} \right)^{d_v-1-\thgbZ - \thgbO} .
\end{align}
In addition, in order to take into account asymmetric deviations, we introduce a third condition, that is that we would like to find parameters $\thgbZ,\thgbO$ that minimize the gap  
\begin{equation} \label{eq:gap_asym}
|\Pv_0(1) - \Pv_1(0) | 
\end{equation} 
between $\Pv_0(1)$ and $\Pv_1(0)$.  
This third condition aims to reduce the asymmetry that is introduced between $\Pv_x(0)$  and $\Pv_1(0)$ by the deviations, see~\eqref{eq:GB_noise}.  
%
At the end, at each iteration $\ell$, we propose to select the two parameters $\thgbZ$ and $\thgbO$ as follows:
\begin{enumerate}
 \item \textcolor{black}{From parameters $b_{0,\ell-1}$ and  $b_{1,\ell-1}$ retained at iteration $\ell-1$, and from the corresponding probability density functions $P_x^{(\ell-1)} $ and $\tilde{Q}_x^{(\ell-1)}$, we evaluate the probability density functions  $\tilde{Q}_x^{(\ell)}$ for all possible pairs of values $(\thgbZ,\thgbO) \in \{ \lceil \frac{d_v}{2} \rceil ,\cdots, d_v-1 \}^2$. We then compute the two sides of the two inequalities~\eqref{eq:ineq1_param} and~\eqref{eq:ineq2_param} for all pair of values $(\thgbZ,\thgbO)$.  }
 \item We then identify the pairs of parameters $(\thgbZ,\thgbO)$ that satisfy both~\eqref{eq:ineq1_param} and~\eqref{eq:ineq2_param}. 
 \item We retain the pairs of parameters $(\thgbZ,\thgbO)$ that minimize the sum $\thgbZ + \thgbO$. 
 \item If there is more than one pair that achieves the minimum, we select the pair that minimizes $|\Pv_0(1) - \Pv_1(0) |$. 
\end{enumerate}
In our experiments, we show that considering two parameters with the above optimization method allows to obtain better decoding performance than when considering one single parameter. 


\subsection{Asymmetric offset parameters in the quantized MS decoder}\label{eq:asym_offset}
We now consider the quantized MS decoder introduced in Section~\ref{sec:dec_ms}, and propose to use two scaling parameters $\alp_0$, $\alp_1$, and two offset parameters $\offp$ and $\offn$. 
The messages are now initialized by using the two scaling parameters as
\begin{equation}
 \minit = 
 \left\{
    \begin{array}{ll}
           \Q(\alp_0 y) & \mbox{if }  y \geq 0 , \\
         \Q(\alp_1 y) & \mbox{if } y < 0 , \\
        0 & \mbox{otherwise.}
    \end{array}
\right.
\end{equation}
The VN mapping is still given by~\eqref{eq:VNqMS}, but the CN mapping is now
\begin{equation}\label{eq:CNmapping_asym}
\mctov_{d_c}^{(\ell)} = 
\left\{
    \begin{array}{ll}
          \left(\prod_{d=1}^{d_c-1} \text{sgn} \left(  \mvtoc^{(\ell)}_d \right) \right) \max\left( \min_{d} \left| \mvtoc^{(\ell)}_d  \right| - \offp , 0 \right)  & \mbox{if } \prod_{d=1}^{d_c-1} \text{sgn} \left(  \mvtoc^{(\ell)}_d \right) > 0, \\
        \left(\prod_{d=1}^{d_c-1} \text{sgn} \left(  \mvtoc^{(\ell)}_d \right) \right) \max\left( \min_{d} \left| \mvtoc^{(\ell)}_d  \right| - \offn , 0 \right)  & \mbox{if } \prod_{d=1}^{d_c-1} \text{sgn} \left(  \mvtoc^{(\ell)}_d \right)  < 0, \\
        0 & \mbox{otherwise.}
    \end{array}
\right.
\end{equation}
As a result, the scaling parameter now depends on the sign of the output channel value, and the offset now depends on the sign of the output message.

In order to take these asymmetric parameters into account, the DE equations given in Section~\ref{sec:dec_ms}, are modified as follows.
The \textcolor{black}{probability density functions} $\Pinit_x$ of the initial messages can be evaluated from the quantization intervals and from the cumulative distribution function provided in Appendix~\ref{app:CumulAsym}. The \textcolor{black}{probability density functions} $\Pc_x$ of the CN messages are still evaluated from~\eqref{eq:DE_CN_MS0} and~\eqref{eq:DE_CN_MS1}, except that now the expression of the function $\Phic$ is given by~\eqref{eq:CNmapping_asym} and depends on the two offset values. Finally, the probability \textcolor{black}{probability density functions} $\Pv_x$ and $\PvN_x$ of the VN messages without and with deviations do not change and are still evaluated from~\eqref{eq:DE_VN_MS} and~\eqref{DE_VN_MS_Noisy}.

\textcolor{black}{Finally, in order to optimize both offset and scaling parameters, we perform an exhaustive search in order to select the four values of $\alp_0$, $\alp_1$, $\offp$ and $\offn$ that minimize the decoder error probability $\Pe$ evaluated with asymptotic DE for given channel and deviation conditions.}
Simulation results show the gain in performance obtained with these asymmetric parameters.

\section{Numerical results}\label{sec:simu}
This section provides simulation results for the Gallager~B decoder and for the quantized Min-Sum decoder. For both decoders, it gives threshold values obtained from the noisy DE analysis presented in this paper for various channel and asymmetric deviation models. It also compares the decoders finite-length performance predicted from noisy DE by using the method described in Section~\ref{sec:fl_de}, with the performance evaluated from Monte Carlo simulations. Finally, it evaluates the effect of asymmetric parameters introduced in Section~\ref{sec:optim}.

In the Monte-Carlo simulations performed in this section, we do not consider the all-zero codeword. Instead, we generate information sequences at random and perform the encoding with the generator matrix, see~\cite{richardson2001efficient}.

\subsection{Gallager~B decoder}

\fig{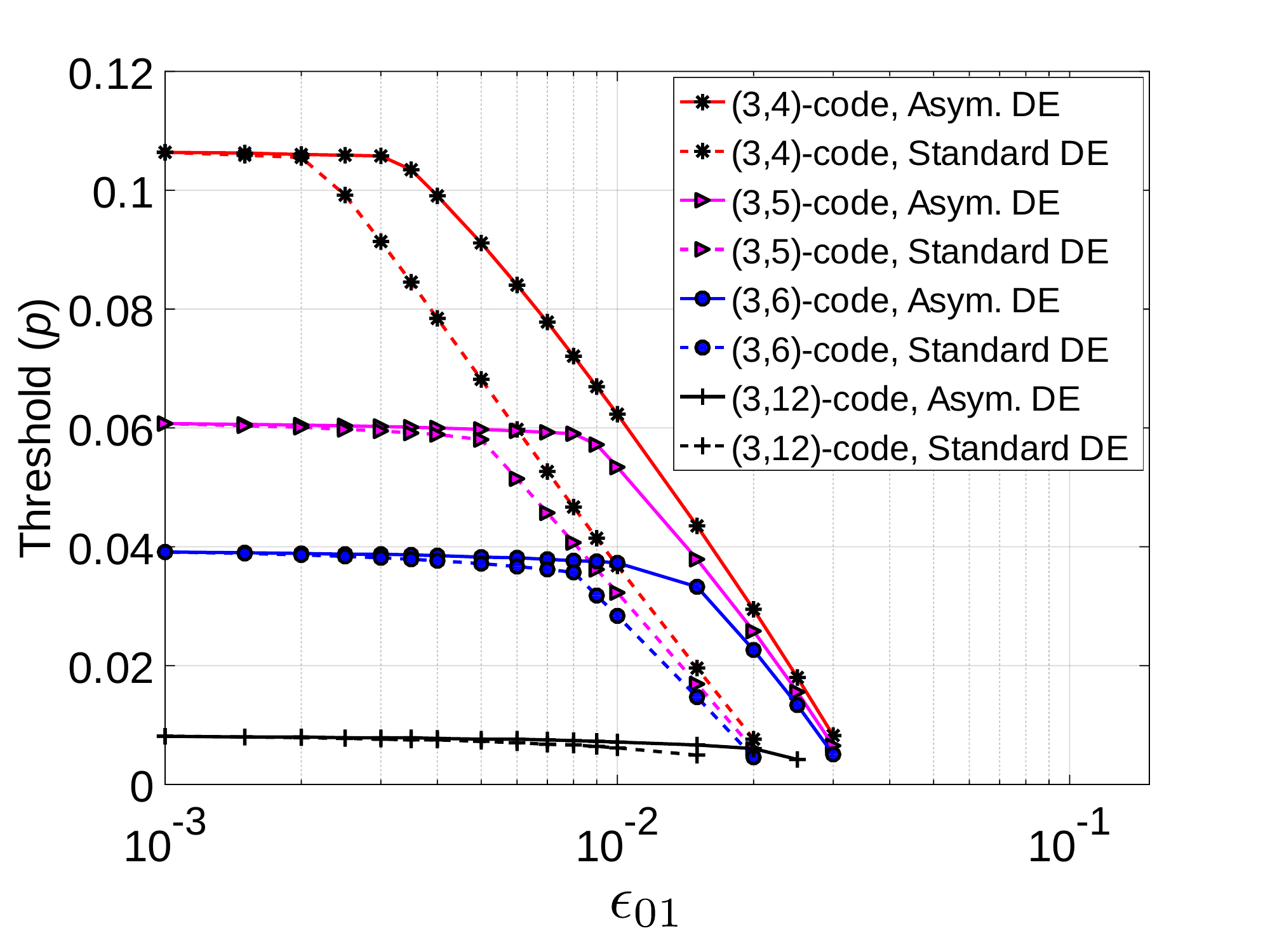}{Regular code ensemble thresholds for $\epn=10^{-3}$  with respect to $\epp$ for the Gallager~B decoder, without the all-zero codeword assumption (standard DE) and with the all-zero codeword assumption (Asym. DE).}{0.6}{t}

In this section, we first provide the code ensemble thresholds obtained for a Gallager~B decoder under asymmetric deviations.
For this decoder, we assume a BSC with crossover probability $p$. 
\textcolor{black}{Following the definition in Section~\ref{sec:th_def}, we measure the threshold as the largest value of $p$ for which $\Pe < 10^{-3}$, where $\Pe$ is expressed in~\eqref{eq:error_prob_GB} with respect to the probability density functions $\PvN_x$ obtained by DE, and where $p$ is found using a binary search.}
We consider four regular codes with $d_v=3$ and $d_c = 4,5,6,12$, respectively, and parameters $L=200$ iterations, $\thgb=2$, and $\epn=10^{-3}$. 
From the DE equations provided in Section~\ref{sec:gb_dec}, we calculate the thresholds for the four considered codes. 
In Figure~\ref{fig:res_gb}, we show the thresholds for different values of $\epp$, obtained with the asymmetric DE presented in this paper, and also the thresholds obtained from standard noisy DE~\cite{varshney2011performance} performed with the all-zero codeword assumption.
\textcolor{black}{The thresholds for standard noisy DE were evaluated by calculating only the probability density functions $\Pc_0$, $\PcN_0$, and $\Pv_0$ in~\eqref{eq:CN_pmf_GB},~\eqref{eq:GB_noise},~\eqref{eq:VN_pmf_GB}, respectively, and $\Pc_0$ was calculated by considering that $\Pv_0 = 1- \Pv_1$, as in standard DE. As a result, the curves for standard DE were evaluated for the same values $\epsilon_{01}\neq \epsilon_{10}$ as in asymmetric DE.  }
As expected, in all the considered cases, we observe from Figure~\ref{fig:res_gb} that the thresholds decrease with $\epp$. 
\textcolor{black}{In addition, we observe that the thresholds obtained under the all-zero codeword assumption differ from the thresholds obtained without this assumption, which shows the need to remove the all-zero codeword assumption in DE under asymmetric deviations.} Nonetheless, for small values of $\epp$, the two threshold values are the same. This comes from the fact that for a small amount of deviations, the noisy thresholds become equal to the noiseless threshold.

\fig{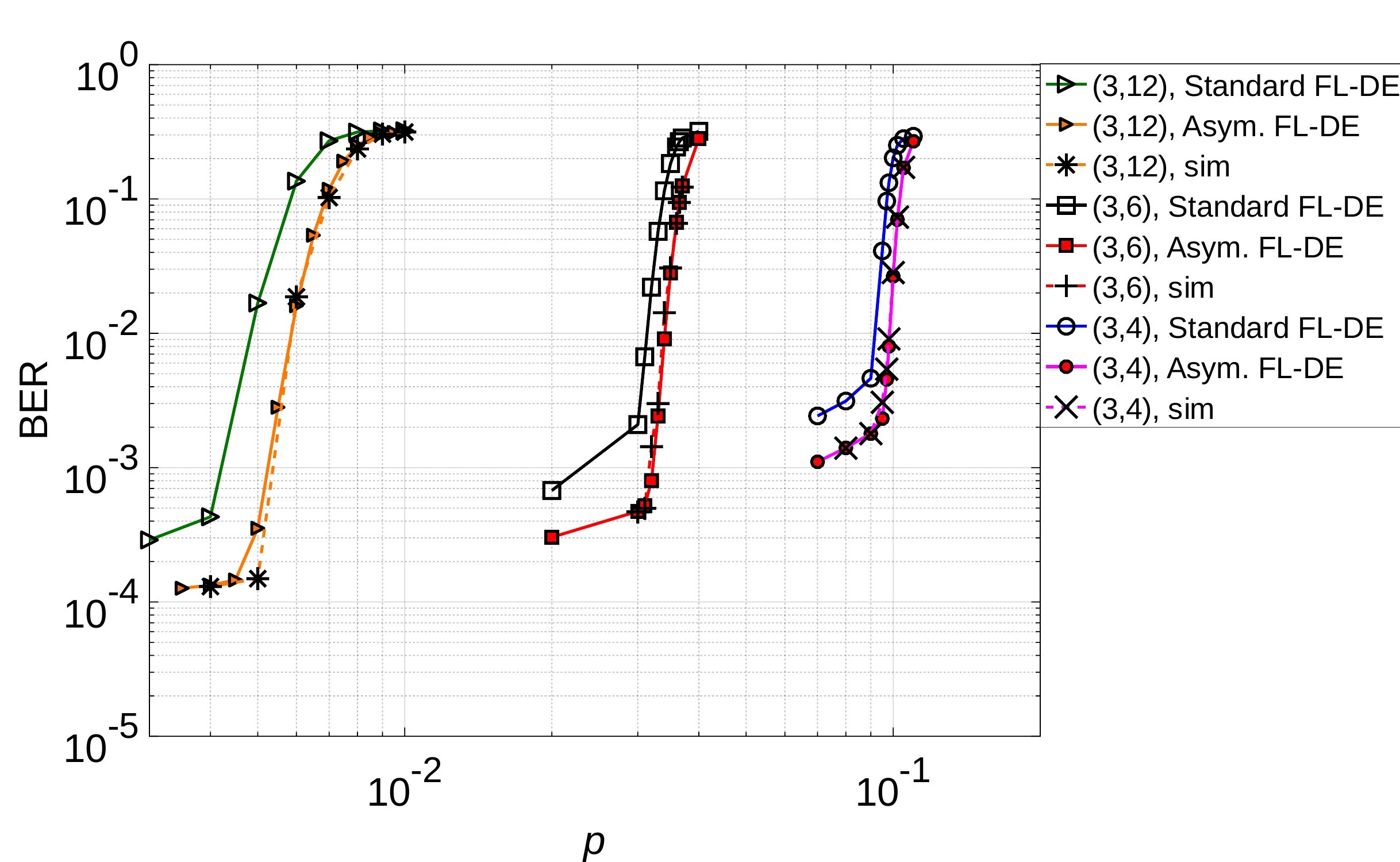}{For the Gallager-B decoder, comparison of BER measured from Monte-Carlo simulations and evaluated with the finite-length DE-based method (FL-DE), with and without the all-zero codeword assumption, for (3,4), (3,6), \textcolor{black}{and (3,12)} regular codes \textcolor{black}{of length $N=10000$, with $\epsilon_{01} = 10^{-2}$, $\epsilon_{10}=10^{-4}$}. For each considered setup, the curves of Monte Carlo simulations and of FL-DE without the all-zero codeword assumption are superimposed.  }{0.7}{t}

Then, in order to verify the accuracy of the proposed asymmetric DE, we use the method described in Section~\ref{sec:fl_de} that allows to predict the decoder performance at finite-length from DE.  
\textcolor{black}{We construct one regular $(3,4)$-code,  one regular $(3,6)$-code, and one regular $(3,12)$-code, both of length $N=10000$, with a Progressive-Edge-Growth (PEG) algorithm~\cite{hu2005regular}. The $(3,4)$ and $(3,6)$ code both have girth $10$, while the $(3,12)$-code has girth $6$}.
We then fix deviation parameters $\epsilon_{01} = 10^{-2}$, $\epsilon_{10}=10^{-4}$. We evaluate the Bit Error Rate (BER) performance of the two codes from Monte Carlo simulations, and compare the obtained BERs to the ones predicted by the finite-length DE-based method described in Section~\ref{sec:fl_de}.
We apply the method of Section~\ref{sec:fl_de} with and without the all-zero codeword assumption.
Figure~\ref{fig:fl_gb} shows the obtained BERs with respect to the crossover parameter $p$. We see that our DE analysis without all-zero codeword assumption accurately predicts the decoder performance of the finite-length codes. On the contrary, we observe a gap between the performance predicted with the all-zero codeword assumption, and the BERs obtained from Monte Carlo simulations. This shows the accuracy and the interest of the method proposed in this paper. 
\textcolor{black}{Finally, Figure~\ref{fig:fl_gb} shows that an error floor appears for both codes for low values of $p$. This error floor is due to deviations introduced in the decoder, which DE can capture, and not to short-length effects such as Tanner graph cycles.  }

\begin{figure}[t]
\begin{center}
  \subfigure[~]{ \includegraphics[width=.48\linewidth]{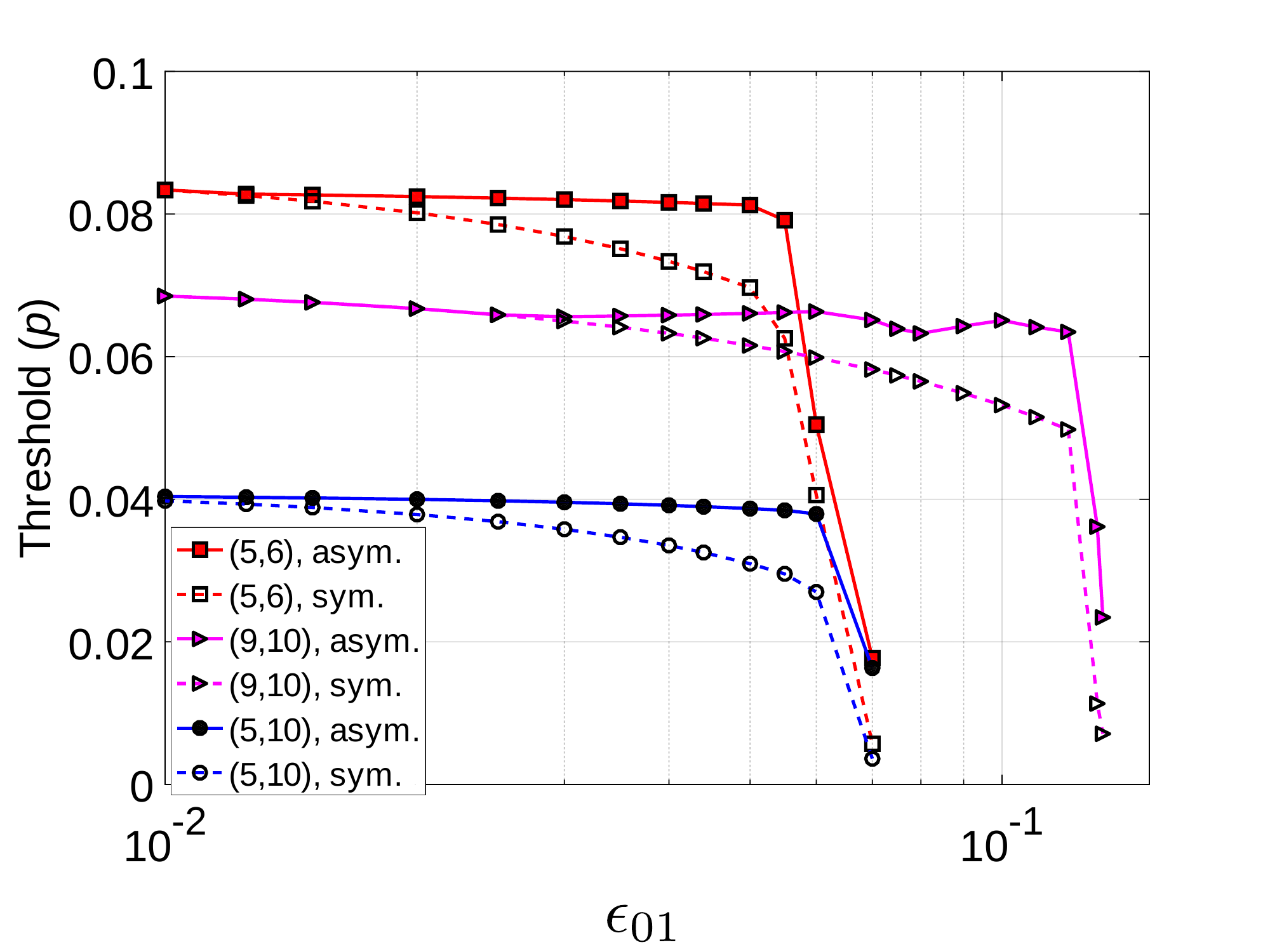}}
  \subfigure[~]{ \includegraphics[width=.48\linewidth]{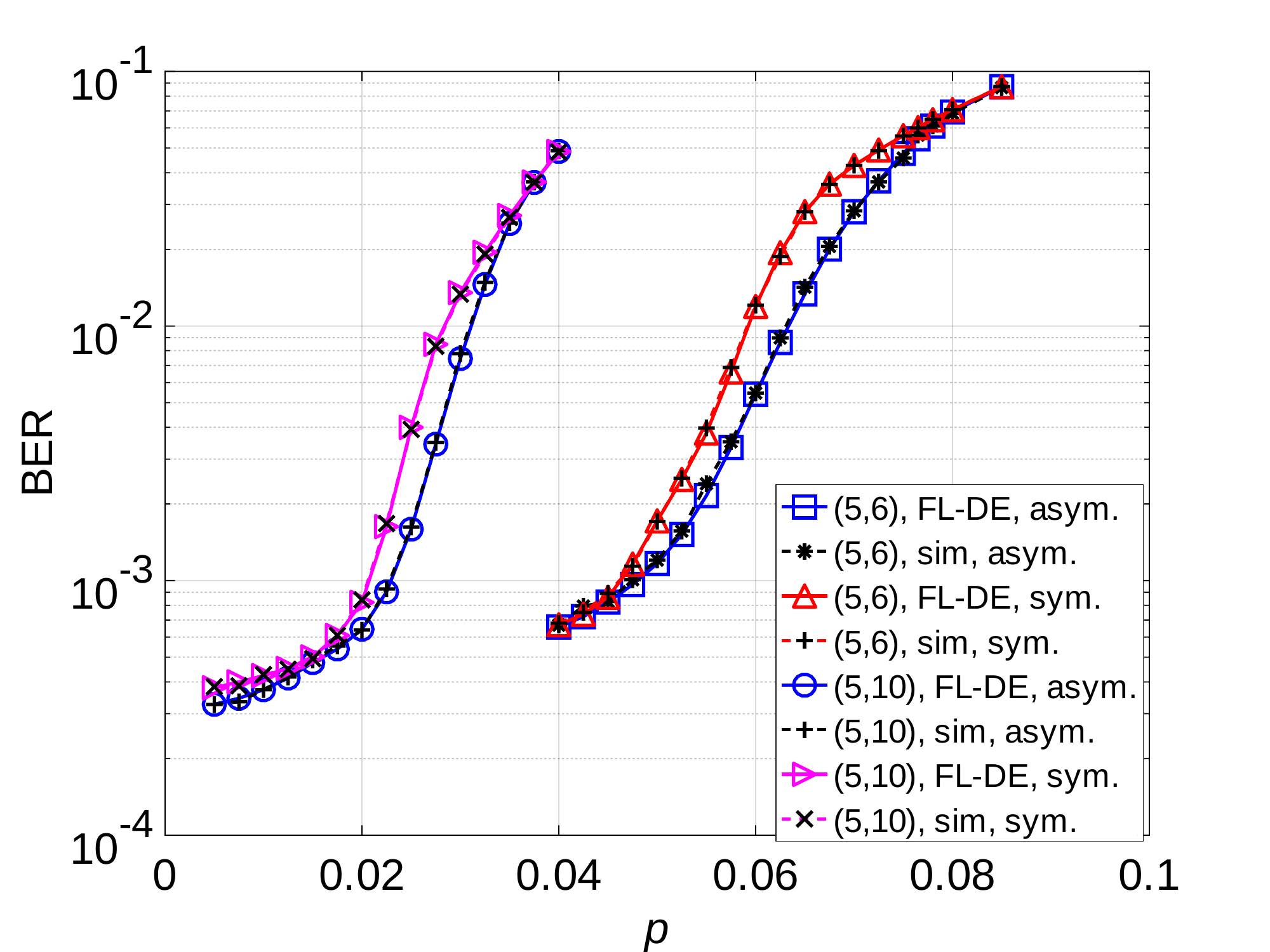}}
\end{center}
\caption{ (a) Threshold optimization with both symmetric and assymmetric Gallager~B decoder parameters, for (5,6), (9,10), and (5,10) regular codes, and for $\epn=10^{-3}$. The $x$-axis label $\epp$ is the deviation parameter from bit value $0$ to bit value $1$. (b) Finite-length performance of $(5,6)$ and $(5,10)$-codes \textcolor{black}{of length $N=10000$} for the Gallager~B decoder with symmetric and asymmetric parameters, with \textcolor{black}{$\epsilon_{01} = 5 \times 10^{-2}$, $\epsilon_{10}=10^{-4}$}. The finite-length performance is measured from Monte-Carlo simulations and evaluated from the finite-length DE-based method (FL-DE) without the all-zero codeword assumption. For each considered setup, the curves of Monte Carlo simulations and of FL-DE are superimposed.}
\label{fig:GB_asym_param}
\end{figure}


Finally, we evaluate the performance of the Gallager~B decoder with asymmetric parameters introduced in Section~\ref{sec:optim}, both from threshold computation and from finite-length performance analysis. For threshold comparison, as before, we set $L=200$ iterations and $\epn=10^{-3}$. We consider three regular $(5,6)$, $(9,10)$, and $(5,10)$ codes. For each code, we consider two Gallager~B decoders: the first one with a symmetric parameter $\thgb$ (optimized with the method of~\cite{richardson2001capacity}), and the second one with two asymmetric parameters $\thgbZ$ and $\thgbO$ (optimized with the method of Section~\ref{sec:optim}). Figure~\ref{fig:GB_asym_param} (a) shows the thresholds obtained for the two methods for various values of $\epsilon_{01}$. We observe that asymmetric parameters improve the decoder thresholds compared to symmetric parameters. 


We also confirm these results on finite-length simulations. For the $(5,6)$ and the $(5,10)$ code, we compare the BER performance obtained with both symmetric and asymmetric decoder parameters. In both cases, we predict the BER from the finite-length DE-based method of Section~\ref{sec:fl_de}, and evaluate the BER from Monte Carlo simulations on codes of length $N=10000$ and girth $8$. The results are shown in Figure~\ref{fig:GB_asym_param} (b) for $\epsilon_{01} = 5 \times 10^{-2}$, $\epsilon_{10}=10^{-4}$, and $L=5$ iterations. We observe that asymmetric parameters clearly improve the BER performance of the Gallager~B decoder under asymmetric deviations.  We also see that the method of Section~\ref{sec:fl_de} accurately predicts the BER performance at finite length, which validates the asymmetric DE analysis introduced in this paper.

\subsection{Quantized Min-Sum decoder}

\begin{figure}[t]
\begin{center}
  \subfigure[~]{ \includegraphics[width=.48\linewidth]{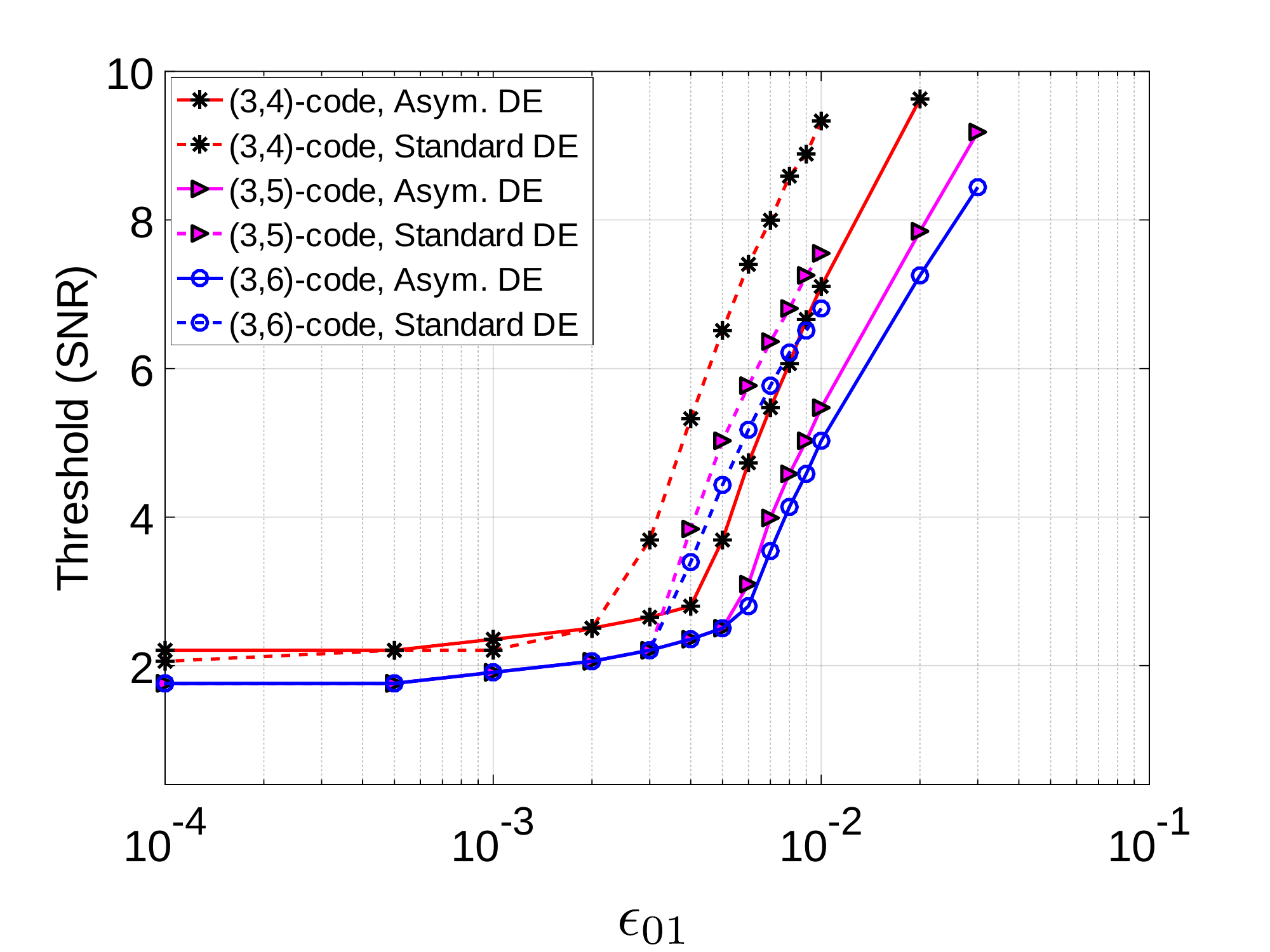}}
  \subfigure[~]{ \includegraphics[width=.48\linewidth]{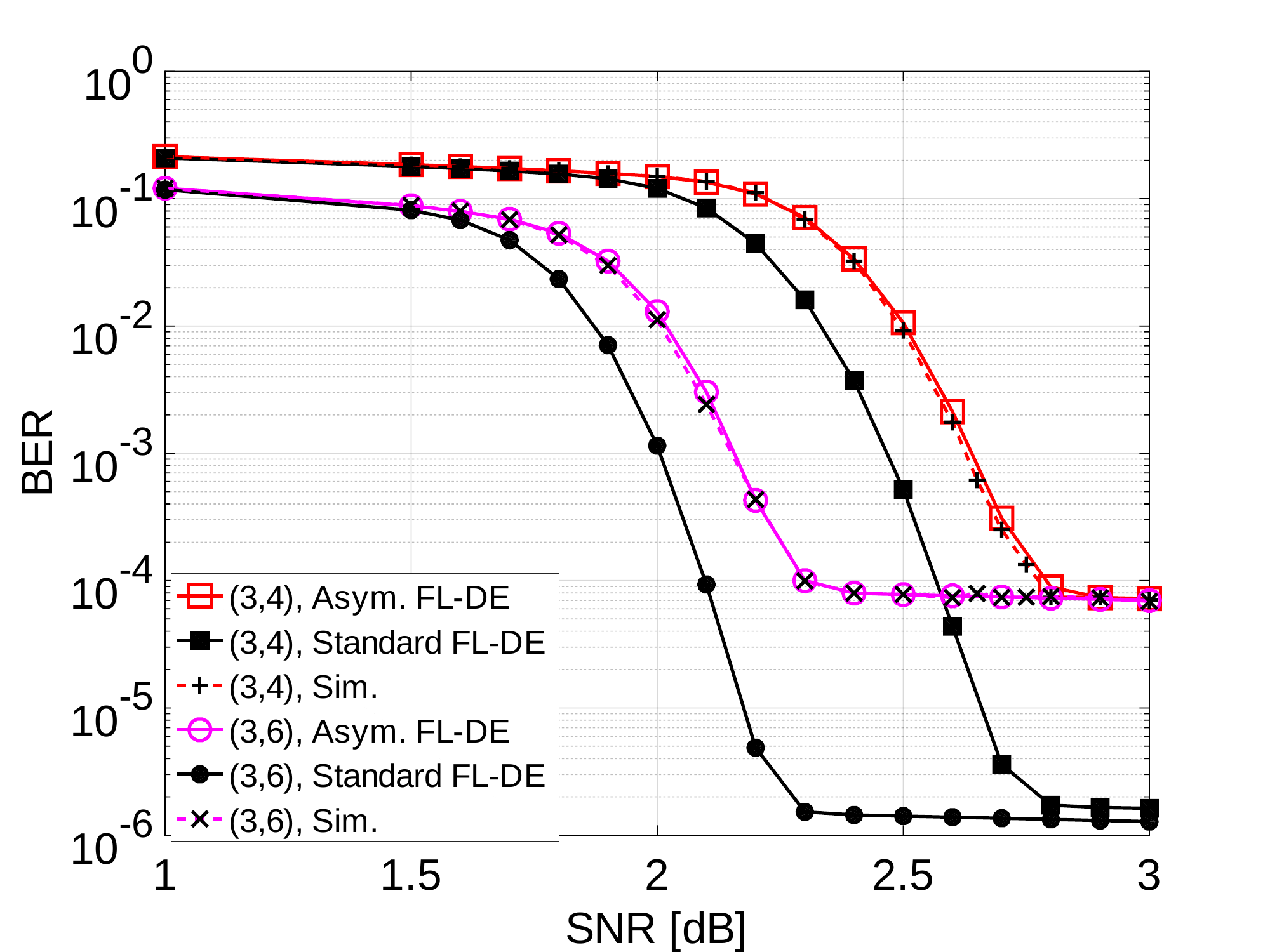}}
\end{center}
\caption{ (a) Regular code ensemble thresholds with respect to $\epp$ for the quantized Min-Sum decoder, with and without the all-zero codeword assumption, and with $\epn=10^{-3}$. The $x$-axis label $\epp$ is the deviation parameter from bit value $0$ to bit value $1$. (b) For the quantized Min-Sum decoder, comparison of BER measured from Monte-Carlo simulations and evaluated with the finite-length DE-based method (FL-DE), with and without the all-zero codeword assumption, for (3,4) and (3,6) regular codes \textcolor{black}{of length $N=10000$, with $\epp = 5\times 10^{-4}$, $\epn = 10^{-3}$}. For each considered setup, the curves of Monte Carlo simulations and of FL-DE without the all-zero codeword assumption are superimposed.  }
\label{fig:MS_asym}
\end{figure}

We now  consider the quantized Min-Sum decoder with the asymmetric deviation model described in Section~\ref{sec:dec_ms}. We consider an AWGN channel with variance $\sigma^2$ and normalized Signal-to-Noise Ratio (SNR) $\text{snr} = 10\log_{10}\left(\frac{1}{2R\sigma^2}\right)$ in dB. In this case, the code ensemble threshold is defined as the smallest value of $\text{snr}$ for which $\Pe < 10^{-3}$ for a large enough $\ell$.
For the decoder parameters, we fix $L=100$ iterations, $q=7$, $\Delta=0.25$, $\alp=1$, $\off=0$, and $\epn=10^{-3}$. We also assume a sign-magnitude binary representation for the messages exchanged in the decoder.
From the DE equations provided in Section~\ref{sec:dec_ms}, we calculate the thresholds with respect to the parameter $\epn$ for four regular codes with VN degree $d_v=3$ and with CN degrees $d_c = 4,5,6,12$, respectively. 
Figure~\ref{fig:MS_asym} (a) shows the thresholds obtained with respect to $\epp$ with standard noisy-DE~\cite{varshney2011performance} with the all-zero codeword assumption, and with the asymmetric noisy-DE analysis introduced in this paper.
\textcolor{black}{The thresholds for standard noisy DE were evaluated by calculating only the probability density functions $\Pc_0$, $\PcN_0$, and $\Pv_0$ in~\eqref{eq:DE_VN_MS},~\eqref{DE_VN_MS_Noisy},~\eqref{eq:DE_CN_MS0} and by considering that $\Pv_0 = 1- \Pv_1$.} 
As  expected, the code ensemble thresholds increase with $\epp$. In addition, the thresholds obtained with standard DE differ from thresholds obtained with our DE analysis, although for small values of $\epp$, the two threshold values are the same.


We then evaluate the finite-length performance of the quantized Min-Sum decoder over one $(3,4)$ regular code and one $(3,6)$ regular code. For the decoder, we set parameters  $L=50$, $q=7$, $\Delta=0.25$, $\alp=1$, $\off=0$, $\epp = 5\times 10^{-4}$, $\epn = 10^{-3}$. For each considered code, we evaluate the BER from Monte-Carlo simulations realized on codes of length $N=10000$ and girth $10$ constructed from the PEG algorithm. We compare the obtained BER values with the performance predicted by the finite-length DE-based method of Section~\ref{sec:fl_de}, evaluated with and with standard DE (with the all-zero codeword assumption) and with our method (without the all-zero codeword assumption) . The results are shown in Figure~\ref{fig:MS_asym} (b). We observe that our method accurately predicts the decoders performance evaluated with Monte-Carlo simulations, while the curve obtained from standard DE shows a gap with the simulations. \textcolor{black}{These results confirm that the DE method introduced in this paper accurately predicts the performance of quantized Min-Sum decoders with asymmetric deviations, while standard Density Evolution with the all-zero codeword assumption is not accurate}. \textcolor{black}{Finally, as for the Gallager B decoder, we observe an error floor at high SNR values, which is due to the deviations in the decoder.}

\fig{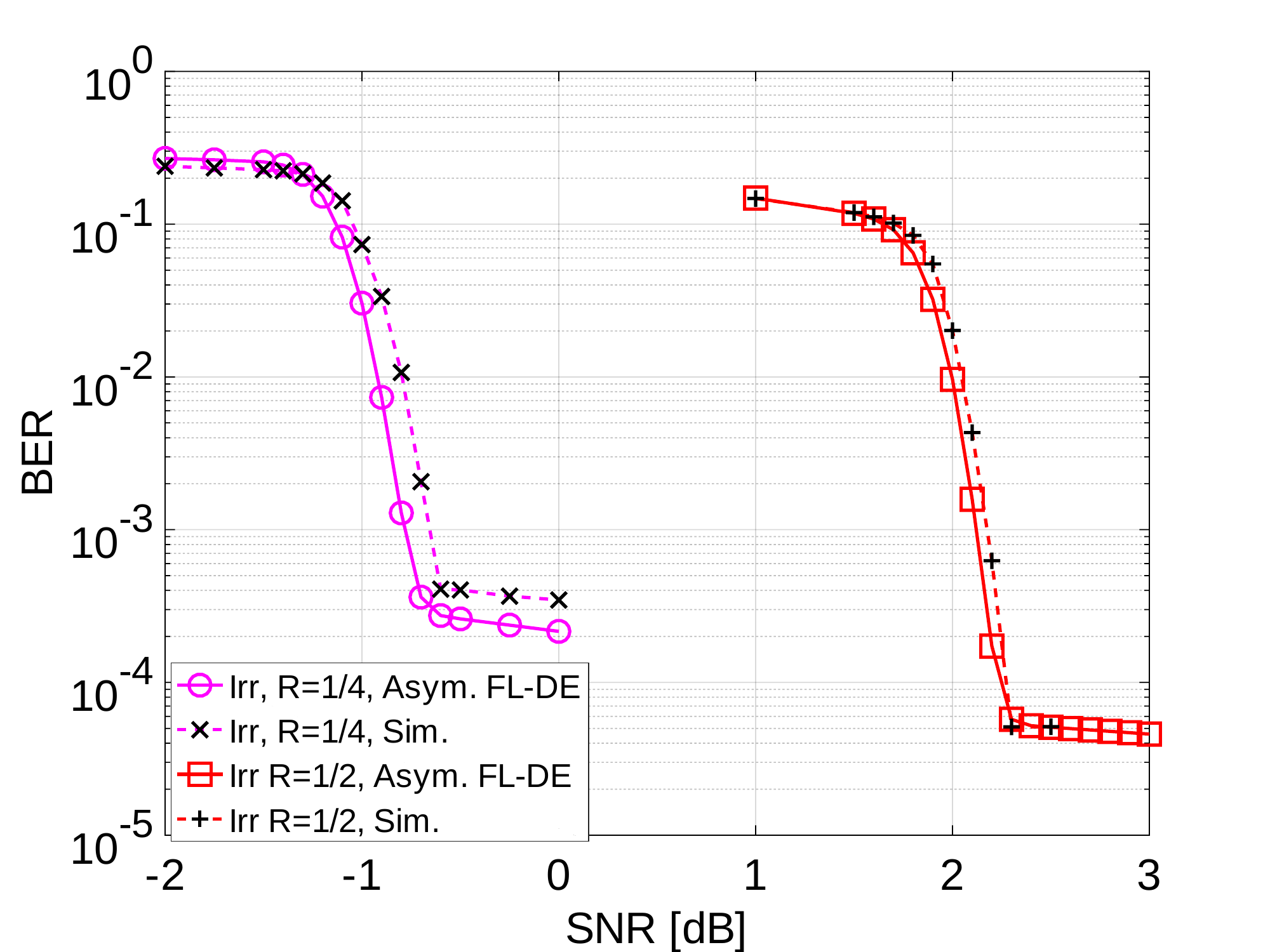}{For the quantized Min-Sum decoder, comparison of BER measured from Monte-Carlo simulations and evaluated with the finite-length DE-based method (FL-DE) with parameters $\epp = 5\times 10^{-4}$, $\epn = 10^{-3}$ for two irregular codes: one with rate $1/4$ and length $N=8000$ and one with rate $1/2$ and length $N=9972$.  }{0.55}{t}

\textcolor{black}{We also evaluate the finite-length performance of the quantized Min-Sum decoder over two irregular LDPC codes, both constructed with the PEG algorithm. The first irregular LDPC code has rate $1/2$, codeword length $N=9972$, and degree distributions $\lambda_1(x) = 0.7857x^2 + 0.2143x^8$, $\rho_1(x)=x^6$~\cite{mackay2009david}. The second irregular code has rate $1/4$, codeword length $N=8000$, and degree distributions $\lambda_2(x) = 0.1688x + 0.1624x^2+0.1313x^8+0.2264x^9+0.007575x^{10}+0.2737x^{14}+0.02986x^{15}$, $\rho_2(x)= 0.19x^5 + 0.81x^6$. For the decoder, we set parameters  $L=50$, $q=5$, $\Delta=0.5$, $\alp=1$, $\off=0$, $\epp = 5\times 10^{-4}$, $\epn = 10^{-3}$. For each of the two irregular codes, Figure~\ref{fig:fl_ms_irreg} compares the BER performance evaluated with Monte-Carlo simulations, with the BER performance predicted by the finite length DE method. This Figure confirms that the asymmetric DE method developed in this paper also allows to consider irregular LDPC codes. The gap between the curves obtained from Monte-Carlo simulations and the curves obtained by finite-length DE are slightly increased compared to regular codes considered in Figure~\ref{fig:MS_asym} (b). This is probably due to the fact that the considered irregular codes are more dense, and hence contain more short cycles, than the regular codes.   }

\begin{figure}[t]
\begin{center}
  \subfigure[~]{ \includegraphics[width=.48\linewidth]{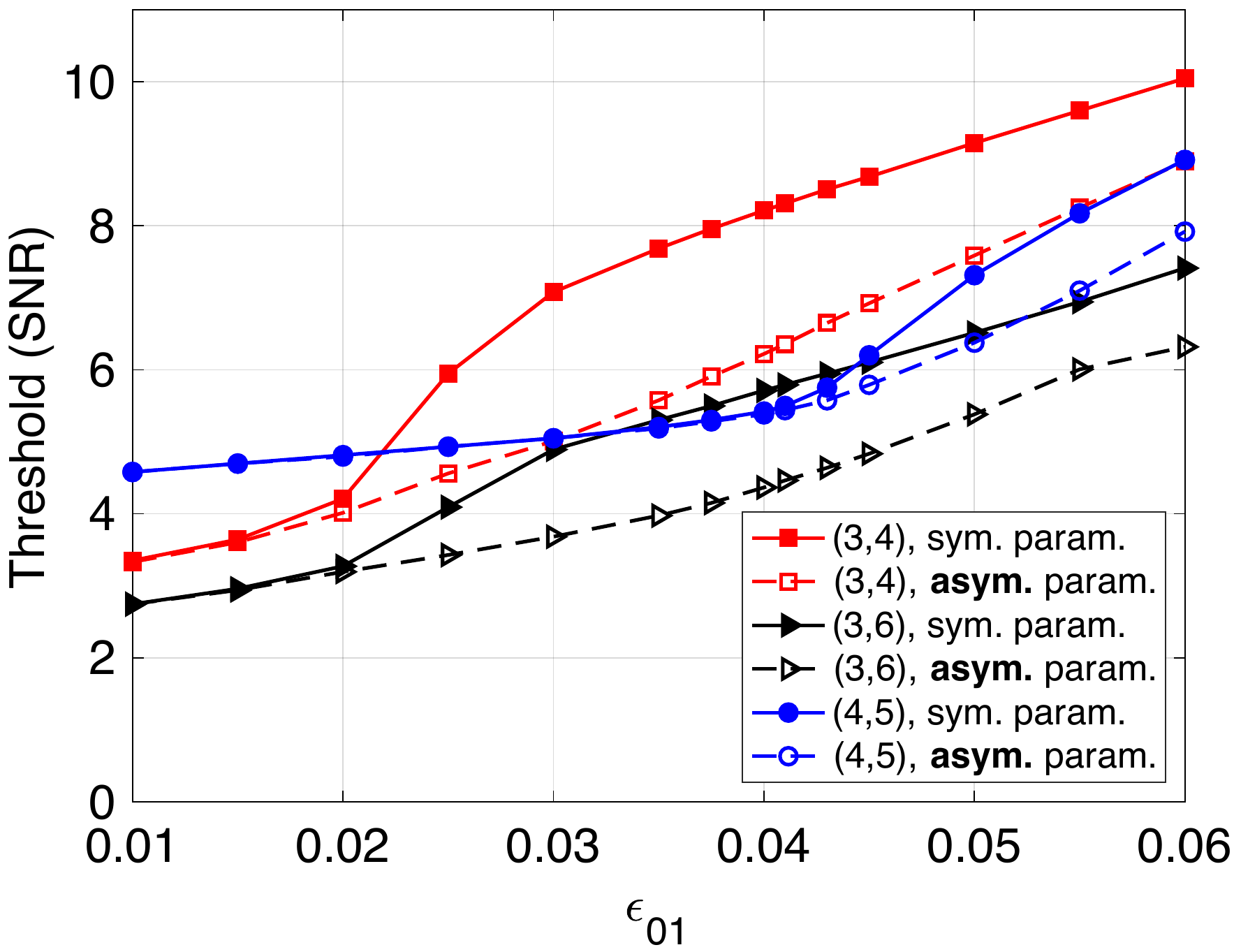}}
  \subfigure[~]{ \includegraphics[width=.48\linewidth]{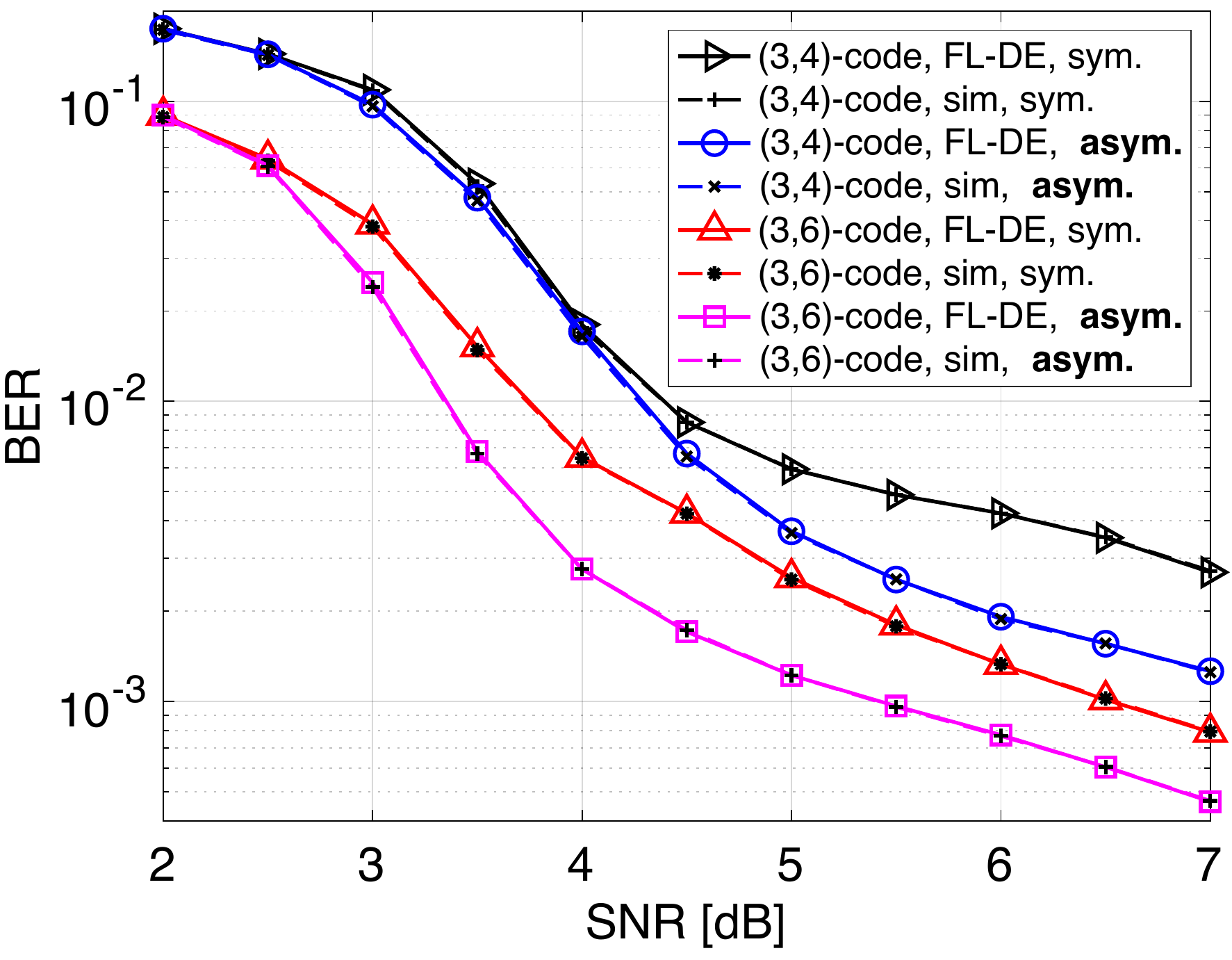}}
\end{center}
\caption{ (a) SNR threshold optimization with symmetric and asymmetric offset and scaling parameters for the quantized Min-Sum decoder, for  $(3,4)$, $(3,6)$, and $(4,5)$-codes, and for $\epn=10^{-5}$. The $x$-axis label $\epp$ is the deviation parameter from bit value $0$ to bit value $1$. (b) Comparison of BER with respect to SNR for the quantized Min-Sum decoder, for symmetric and asymmetric decoder parameters, \textcolor{black}{for codes of length $N=10000$ with $\epn=10^{-5}$, and $\epp=5 \times 10^{-2}$}. For each considered setup, plain curves (performance evaluated from FL-DE) and dashed curves (performance evaluated from Monte Carlo simulations) are superimposed.}
\label{fig:MS_asym_param}
\end{figure}


\begin{table}[t]
\centering
\caption{Optimized symmetric and asymmetric parameter values for the quantized Min-Sum decoder, for $(3,4)$, $(3,6)$, and $(4,5)$-codes, for $\epn=10^{-5}$. }\label{tab:parameters_MS}
\begin{tabular}{|c|c||c|c|c||c|c|c|c|c|}
 \hline
 \multicolumn{2}{|l||}{ } &  \multicolumn{3}{c||}{ Symmetric parameters } & \multicolumn{5}{c|}{ Asymmetric parameters } \\
 \hline
 code & $\epp$ & $\gamma$ & $\lambda$ & Threshold & $\gamma_0$ & $\gamma_1$ & $\lambda_0$ & $\lambda_1$ & Threshold \\
 \hline \hline
 (3,4) & 0.01 & 0.85 & 0  & 3.45 dB  & 0.90  & 0.80  & 0  & 0  & 3.30 dB \\
  & 0.03 & 0.55  & 1  & 7.08 dB & 1.0  & 0.25  & 1 & 0 & 5.01 dB \\
 \hline \hline
 (3,6) & 0.01 & 0.70 & 0 & 2.75 dB & 0.75 & 0.7  & 0  & 0  & 2.74 dB\\
   & 0.03 & 0.60 & 1  & 4.89 dB & 1.0 & 0.35  & 1  & 0  & 3.68 dB \\
 \hline \hline
 (4,5) & 0.01 & 0.95 & 0  & 4.58 dB  & 0.95  & 0.90  & 0 & 0  & 4.48 dB\\
  & 0.05 & 1.0  & 0   & 7.31 dB  & 1.0  & 0.25  & 1  & 0  & 6.38 dB\\
 \hline
\end{tabular}
\vspace{0.3cm} 
\end{table}

We now evaluate the performance of the quantized Min-Sum decoder with asymmetric scaling and offset parameters introduced in Section~\ref{eq:asym_offset}. We consider regular $(3,4)$, $(3,6)$, and $(4,5)$-codes, and decoder parameters given by $L=10$ iterations, $q=4$, $\Delta=1$, and $\epn=10^{-5}$.  
In Figure~\ref{fig:MS_asym_param} (a), we show the SNR thresholds with respect to the noise parameter $\epsilon_{01}$, obtained by considering symmetric and asymmetric parameters. We observe that asymmetric parameters improve the decoder performance compared to symmetric parameters. In our simulations, we observed that both the scaling parameter and the offset parameter need to be asymmetric. For instance, symmetric scaling parameter with asymmetric offset do not improve much the performance compared to the fully symmetric setup. 
In addition, Table~\ref{tab:parameters_MS} provides the optimized values of the symmetric and asymmetric parameters as well as the corresponding thresholds, for two values of $\epn$ for the three considered codes. The values in the Table confirm that asymmetric parameters allow to improve the decoder thresholds.


At the end, we confirm the interest of asymmetric quantized Min-Sum parameters on finite-length simulations. We consider decoder parameters given by $L=10$ iterations, $q=4$, $\Delta=1$, $\epn=10^{-5}$, and $\epp=5 \times 10^{-2}$. For $(3,4)$ and the $(3,6)$ regular codes, we compare the BER performance obtained with optimized symmetric and asymmetric decoder parameters. In both cases, we predict the BER from the finite-length DE-based method of Section~\ref{sec:fl_de}, and evaluate the BER performance from Monte Carlo simulations on codes of length $N=10000$ and girth $8$ obtained from a PEG algorithm. The results are shown in Figure~\ref{fig:MS_asym_param} (b), and confirm that asymmetric parameters clearly improve the BER performance of the quantized Min-Sum decoder under asymmetric deviations.  We also see that the method of Section~\ref{sec:fl_de} accurately predicts the BER performance at finite length, which validates the asymmetric DE analysis introduced in this paper.

\section{Conclusion}
We considered noisy LDPC decoders under asymmetric deviation models and derived the noisy DE equations without the all-zero codeword assumption for three noisy decoders: the BP decoder, the Gallager~B decoder, and the quantized offset MS decoder.
We then proposed to compensate the effects of asymmetric deviations by introducing asymmetric parameters in the Gallager~B and quantized MS decoders. Numerical simulations confirmed both the accuracy of the proposed DE method, and the performance improvement provided by the use of asymmetric parameters.
Future work will target the optimization of irregular degree distributions and protographs under asymmetric deviation models.

\section*{Acknowledgement}
The authors were supported by the grant ANR-17-CE40-0020 of the French National Research Agency ANR (project EF-FECtive). 

\appendices

\section{CN message \textcolor{black}{probability density function} in the BP decoder}\label{app:proofCN}
 We derive the CN output message \textcolor{black}{probability density function}~\eqref{eq:probaCN_BP} for the BP decoder. At the CN, the output message is denoted $\mctov$ and the input messages are denoted $\mvtoc_{i}$. We first consider the case where the VN that receives message $\mctov$ has value $x=0$. Denote by $x_1,\cdots, x_{d_c-1}$ the bit values of the VNs that sent messages $\mvtoc_1,\cdots, \mvtoc_{d_c-1}$. Denote by $V$ the random variable that represents the number of values $1$ among $x_1,\cdots,x_{d_c-1}$.
 By marginalization with respect to $V$, we have
 \begin{equation}
  \Pc_0(\mctov) = \Pr(\mctov|x=0) = \sum_{v=0}^{d_c-1} \Pr(V=v|x=0) \Pr(\mctov|x=0, V=v). 
 \end{equation}
If $v$ is odd, then $\Pr(V=v|X=0) = 0$, since $x + \sum_{i=1}^{d_c-1} x_i = 0$. If $v$ is even, then $$\Pr(V=v|X=0) =  \binom{d_c-1}{v} \left(\frac{1}{2}\right)^{d_c-2}. $$ In addition, $$\Pr(\mctov|x=0, V=v) = \Gamma^{-1} \left( \Gamma\left( \PvN_0 \right)^{\otimes(d_c-1-v)} \otimes \Gamma\left( \PvN_1 \right)^{\otimes v} \right),$$ since $(x_1,\cdots,x_{d_c-1})$ contains $v$ values $1$ and $(d_c-1-v)$ values $0$. Combining these expressions gives $\Pc_0(\mctov)$ in~\eqref{eq:probaCN_BP}. 
The derivation of $\Pc_1(\mctov)$ follows the same process.
 

\section{\textcolor{black}{Probability density function} of initial messages in the quantized Min-Sum decoder with asymmetric parameters}\label{app:CumulAsym}
In this part, we consider an AWGN channel with variance $\sigma^2$.
In order to compute the probability density functions $\Pinit_x$ of initial messages in the quantized Min-Sum decoder with asymmetric parameters, we need to determine the cumulative distribution function of a random variable $ z = \frac{2\gamma(y)y}{\sigma^2} $, where $\gamma(y) = \gamma_0$ if $y\geq 0$, and   $\gamma(y) = \gamma_1$ if $y < 0$. 
We denote by $\mathbb{F}_{z|x}(t) = \mathbb{P}(z \leq t | x)$ the cumulative distribution function of the random variable $z$ conditioned on $x\in \{0,1\}$.

The cumulative distribution function $\mathbb{F}_{z|x}(t)$ can be expressed as
\begin{align}
 \mathbb{F}_{z|x}(t) & = \mathbb{P}\left(  \frac{2\gamma(y)y}{\sigma^2} \leq t |x \right) \\
        & = \mathbb{P}\left( \left( \frac{2\gamma(y)y}{\sigma^2} \leq t\right) \cap \left(y\geq 0\right) | x \right) + \mathbb{P}\left(  \left( \frac{2\gamma(y)y}{\sigma^2} \leq t\right) \cap \left(y\leq 0\right) |x \right)
\end{align}
by the law of total probability.
We then treat separately the case where $t < 0$ and the case $t \geq 0$, starting with the first case.

Since $\gamma(y) > 0$, the condition $t<0$ implies that $y < 0$. As a result, when $t<0$, the cumulative distribution function $\mathbb{F}_{z|x}(t)$ is given by
 $\mathbb{F}_{z|x}(t) = \mathbb{P}\left( \frac{2\gamma_1 y}{\sigma^2} | x \right).$
We denote $\mu_1(x)  = \frac{2\gamma_1 (1-2x)}{\sigma^2}$ and $\sigma_1^2 = \frac{4\gamma_1^2}{\sigma^2}$. Then, the random variable $\frac{2\gamma_1 y}{\sigma^2}$ follows a Gaussian distribution with mean $\mu_1(x)$ and variance $\sigma_1^2$. As a result,
\begin{equation}
 \forall t < 0 ~ , ~ \mathbb{F}_{z|x}(t) = \frac{1}{2} + \frac{1}{2}\text{erf}\left( \frac{t-\mu_1(x)}{\sqrt{2} \sigma_1} \right)
\end{equation}
where $\text{erf}$ is the error function of the Gaussian distribution.

Now, for $t \geq 0$, the cumulative distribution function $\mathbb{F}_{z|x}(t)$ is given by
\begin{equation}
 \mathbb{F}_{z|x}(t) = \mathbb{P}\left( \frac{2\gamma(y)y}{\sigma^2} \leq 0 | x \right) + \mathbb{P}\left(  0 \leq \frac{2\gamma(y)y}{\sigma^2} \leq t|x \right) .
\end{equation}
We denote  $\mu_0(x)  = \frac{2\gamma_0 (1-2x)}{\sigma^2}$ and $\sigma_0^2 = \frac{4\gamma_0^2}{\sigma^2}$. Then, the random variable $\frac{2\gamma_0 y}{\sigma^2}$ follows a Gaussian distribution with mean $\mu_0(x)$ and variance $\sigma_0^2$. As a result,
\begin{equation}
 \forall t \geq 0 ~ , ~ \mathbb{F}_{z|x}(t) = \frac{1}{2} + \frac{1}{2}\text{erf}\left( -\frac{\mu_1(x)}{\sqrt{2} \sigma_1} \right) + \frac{1}{2}\text{erf}\left( \frac{\mu_0(x)}{\sqrt{2} \sigma_0} \right) + \frac{1}{2}\text{erf}\left( \frac{t-\mu_0(x)}{\sqrt{2} \sigma_0} \right) .
\end{equation}

\bibliography{biblio}

\begin{thebibliography}{10}
\providecommand{\url}[1]{#1}
\csname url@samestyle\endcsname
\providecommand{\newblock}{\relax}
\providecommand{\bibinfo}[2]{#2}
\providecommand{\BIBentrySTDinterwordspacing}{\spaceskip=0pt\relax}
\providecommand{\BIBentryALTinterwordstretchfactor}{4}
\providecommand{\BIBentryALTinterwordspacing}{\spaceskip=\fontdimen2\font plus
\BIBentryALTinterwordstretchfactor\fontdimen3\font minus
  \fontdimen4\font\relax}
\providecommand{\BIBforeignlanguage}[2]{{%
\expandafter\ifx\csname l@#1\endcsname\relax
\typeout{** WARNING: IEEEtran.bst: No hyphenation pattern has been}%
\typeout{** loaded for the language `#1'. Using the pattern for}%
\typeout{** the default language instead.}%
\else
\language=\csname l@#1\endcsname
\fi
#2}}
\providecommand{\BIBdecl}{\relax}
\BIBdecl

\bibitem{gupta:2013}
P.~{Gupta}, Y.~{Agarwal}, L.~{Dolecek}, N.~{Dutt}, R.~K. {Gupta}, R.~{Kumar},
  S.~{Mitra}, A.~{Nicolau}, T.~S. {Rosing}, M.~B. {Srivastava}, S.~{Swanson},
  and D.~{Sylvester}, ``Underdesigned and opportunistic computing in presence
  of hardware variability,'' \emph{IEEE Transactions on Computer-Aided Design
  of Integrated Circuits and Systems}, vol.~32, no.~1, pp. 8--23, Jan 2013.

\bibitem{yang2017computing}
Y.~Yang, P.~Grover, and S.~Kar, ``Computing linear transformations with
  unreliable components,'' \emph{IEEE Transactions on Information Theory},
  vol.~63, no.~6, pp. 3729--3756, 2017.

\bibitem{yang2016fault}
------, ``Fault-tolerant distributed logistic regression using unreliable
  components,'' in \emph{2016 54th Annual Allerton Conference on Communication,
  Control, and Computing (Allerton)}.\hskip 1em plus 0.5em minus 0.4em\relax
  IEEE, 2016, pp. 940--947.

\bibitem{dupraz2019binary}
E.~Dupraz and L.~R. Varshney, ``Binary recursive estimation on noisy
  hardware,'' in \emph{2019 IEEE International Symposium on Information Theory
  (ISIT)}.\hskip 1em plus 0.5em minus 0.4em\relax IEEE, 2019, pp. 877--881.

\bibitem{hacene2019training}
G.~B. Hacene, F.~Leduc-Primeau, A.~B. Soussia, V.~Gripon, and F.~Gagnon,
  ``Training modern deep neural networks for memory-fault robustness,'' in
  \emph{2019 IEEE International Symposium on Circuits and Systems
  (ISCAS)}.\hskip 1em plus 0.5em minus 0.4em\relax IEEE, 2019, pp. 1--5.

\bibitem{varshney2011performance}
L.~Varshney, ``{Performance of {LDPC} codes under faulty iterative decoding},''
  \emph{IEEE Transactions on Information Theory}, vol.~57, no.~7, pp.
  4427--4444, 2011.

\bibitem{leduc-primeau:2012}
F.~Leduc-Primeau and W.~J. Gross, ``Faulty {Gallager-B} decoding with optimal
  message repetition,'' in \emph{Proc. 50th Allerton Conf. on Communication,
  Control, and Computing}, Oct. 2012.

\bibitem{al2014fault}
O.~Al~Rasheed, P.~Ivani{\v{s}}, and B.~Vasi{\'c}, ``Fault-tolerant
  probabilistic gradient-descent bit flipping decoder,'' \emph{IEEE
  Communications Letters}, vol.~18, no.~9, pp. 1487--1490, 2014.

\bibitem{sundararajan2014noisy}
G.~Sundararajan, C.~Winstead, and E.~Boutillon, ``Noisy gradient descent
  bit-flip decoding for {LDPC} codes,'' \emph{IEEE Transactions on
  Communications}, vol.~62, no.~10, p.~16, 2014.

\bibitem{Huang2014Gallager}
C.-H. Huang, Y.~Li, and L.~Dolecek, ``{Gallager B {LDPC} decoder with transient
  and permanent errors},'' \emph{IEEE Transactions on Communications}, vol.~62,
  no.~1, pp. 15--28, 2014.

\bibitem{ngassa2015density}
C.~K. Ngassa, V.~Savin, E.~Dupraz, and D.~Declercq, ``Density evolution and
  functional threshold for the noisy min-sum decoder,'' \emph{IEEE Transactions
  on Communications}, vol.~63, no.~5, pp. 1497--1509, 2015.

\bibitem{balatsoukas2014density}
A.~Balatsoukas-Stimming and A.~Burg, ``Density evolution for min-sum decoding
  of {LDPC} codes under unreliable message storage,'' \emph{IEEE Communications
  Letters}, vol.~18, no.~5, pp. 849--852, 2014.

\bibitem{leduc2018modeling}
F.~Leduc-Primeau, F.~R. Kschischang, and W.~J. Gross, ``Modeling and energy
  optimization of {LDPC} decoder circuits with timing violations,'' \emph{IEEE
  Transactions on Communications}, vol.~66, no.~3, pp. 932--946, 2018.

\bibitem{Dupraz15Com}
E.~Dupraz, D.~Declercq, B.~Vasi\'c, and V.~Savin, ``Analysis and design of
  finite alphabet iterative decoders robust to faulty hardware,'' \emph{IEEE
  Transactions on Communications}, vol.~63, no.~8, pp. 2797--2809, 2015.

\bibitem{richardson2001capacity}
T.~J. Richardson and R.~L. Urbanke, ``The capacity of low-density parity-check
  codes under message-passing decoding,'' \emph{IEEE Transactions on
  Information Theory}, vol.~47, no.~2, pp. 599--618, 2001.

\bibitem{richardson2001design}
T.~J. Richardson, M.~A. Shokrollahi, and R.~L. Urbanke, ``Design of
  capacity-approaching irregular low-density parity-check codes,'' \emph{IEEE
  Transactions on Information Theory}, vol.~47, no.~2, pp. 619--637, 2001.

\bibitem{wang2005density}
C.-C. Wang, S.~R. Kulkarni, and H.~V. Poor, ``Density evolution for asymmetric
  memoryless channels,'' \emph{IEEE Transactions on Information Theory},
  vol.~51, no.~12, pp. 4216--4236, 2005.

\bibitem{chen2009equivalence}
J.~Chen, D.-k. He, and A.~Jagmohan, ``The equivalence between {Slepian-Wolf}
  coding and channel coding under density evolution,'' \emph{IEEE Transactions
  on Communications}, vol.~57, no.~9, 2009.

\bibitem{Bravo:2019}
E.~V. {Bravo}, A.~{Bonetti}, and A.~{Burg}, ``Data-retention-time
  characterization of gain-cell {eDRAMs} across the design and variations
  space,'' in \emph{2019 IEEE International Symposium on Circuits and Systems
  (ISCAS)}, May 2019.

\bibitem{verma:2008}
N.~{Verma} and A.~P. {Chandrakasan}, ``A 256 kb 65 nm 8t subthreshold sram
  employing sense-amplifier redundancy,'' \emph{IEEE Journal of Solid-State
  Circuits}, vol.~43, no.~1, pp. 141--149, 2008.

\bibitem{nabavi:2016}
M.~{Nabavi}, F.~{Ramezankhani}, and M.~{Shams}, ``Optimum pmos-to-nmos width
  ratio for efficient subthreshold cmos circuits,'' \emph{IEEE Transactions on
  Electron Devices}, vol.~63, no.~3, pp. 916--924, 2016.

\bibitem{nadal:2020a}
J.~{Nadal}, M.~{Fiorentino}, E.~{Dupraz}, and F.~{Leduc-Primeau}, ``A deeply
  pipelined, highly parallel and flexible ldpc decoder,'' in \emph{2020 18th
  IEEE International New Circuits and Systems Conference (NEWCAS)}, 2020, pp.
  263--266.

\bibitem{savin2008self}
V.~Savin, ``Self-corrected min-sum decoding of {LDPC} codes,'' in \emph{2008
  IEEE International Symposium on Information Theory}.\hskip 1em plus 0.5em
  minus 0.4em\relax IEEE, 2008, pp. 146--150.

\bibitem{dupraz2014density}
E.~Dupraz, V.~Savin, and M.~Kieffer, ``Density evolution for the design of
  non-binary low density parity check codes for {Slepian-Wolf} coding,''
  \emph{IEEE Transactions on Communications}, vol.~63, no.~1, pp. 25--36, 2014.

\bibitem{leduc2016finite}
F.~Leduc-Primeau and W.~J. Gross, ``Finite-length quasi-synchronous {LDPC}
  decoders,'' in \emph{9th International Symposium on Turbo Codes and Iterative
  Information Processing (ISTC)}.\hskip 1em plus 0.5em minus 0.4em\relax IEEE,
  2016, pp. 325--329.

\bibitem{friedman2001elements}
J.~Friedman, T.~Hastie, and R.~Tibshirani, \emph{The elements of statistical
  learning}.\hskip 1em plus 0.5em minus 0.4em\relax Springer series in
  statistics New York, 2001, vol.~1, no.~10.

\bibitem{hu2005regular}
X.-Y. Hu, E.~Eleftheriou, and D.-M. Arnold, ``Regular and irregular progressive
  edge-growth {Tanner} graphs,'' \emph{IEEE Transactions on Information
  Theory}, vol.~51, no.~1, pp. 386--398, 2005.

\bibitem{tarighati2015design}
A.~Tarighati, H.~Farhadi, and F.~Lahouti, ``Design of {LDPC} codes robust to
  noisy message-passing decoding,'' \emph{arXiv preprint arXiv:1501.02483},
  2015.

\bibitem{gorgoglione2010optimized}
M.~Gorgoglione, V.~Savin, and D.~Declercq, ``Optimized puncturing distributions
  for irregular non-binary ldpc codes,'' in \emph{2010 International Symposium
  On Information Theory \& Its Applications}.\hskip 1em plus 0.5em minus
  0.4em\relax IEEE, 2010, pp. 400--405.

\bibitem{richardson2001efficient}
T.~J. Richardson and R.~L. Urbanke, ``Efficient encoding of low-density
  parity-check codes,'' \emph{IEEE Transactions on Information Theory},
  vol.~47, no.~2, pp. 638--656, 2001.

\bibitem{mackay2009david}
D.~MacKay, ``David mackay’s gallager code resources,''
  \emph{http://www.inference.phy.cam.ac.uk/mackay/CodesFiles.html}, 2009.

\end{thebibliography}
\bibliographystyle{IEEEtran}

\end{document}